\documentstyle[psfig,epsfig]{mn}

\newif\ifAMStwofonts
\newcommand{\msolar}{\mbox{\,$M_{\odot}$}}

\ifoldfss
  \ifCUPmtlplainloaded \else
    \NewTextAlphabet{textbfit} {cmbxti10} {}
    \NewTextAlphabet{textbfss} {cmssbx10} {}
    \NewMathAlphabet{mathbfit} {cmbxti10} {} % for math mode
    \NewMathAlphabet{mathbfss} {cmssbx10} {} %  "   "    "
  \fi
  \ifAMStwofonts
    \ifCUPmtlplainloaded \else
      \NewSymbolFont{upmath} {eurm10}
      \NewSymbolFont{AMSa} {msam10}
      \NewMathSymbol{\upi}     {0}{upmath}{19}
      \NewMathSymbol{\umu}     {0}{upmath}{16}
      \NewMathSymbol{\upartial}{0}{upmath}{40}
      \NewMathSymbol{\leqslant}{3}{AMSa}{36}
      \NewMathSymbol{\geqslant}{3}{AMSa}{3E}

      \let\leq=\leqslant 
      \let\geq=\geqslant \let\ge=\geqslant
    \fi
  \fi
\fi % End of OFSS

\ifnfssone
  \newmathalphabet{\mathit}
  \addtoversion{normal}{\mathit}{cmr}{m}{it}
  \addtoversion{bold}{\mathit}{cmr}{bx}{it}
  \newmathalphabet{\mathbfit} % math mode version of \textbfit{..}
  \addtoversion{normal}{\mathbfit}{cmr}{bx}{it}
  \addtoversion{bold}{\mathbfit}{cmr}{bx}{it}
  \newmathalphabet{\mathbfss} % math mode version of \textbfss{..}
  \addtoversion{normal}{\mathbfss}{cmss}{bx}{n}
  \addtoversion{bold}{\mathbfss}{cmss}{bx}{n}
  \ifAMStwofonts
    \ifCUPmtlplainloaded \else
      \UseAMStwoboldmath
      \makeatletter
      \new@mathgroup\upmath@group
      \define@mathgroup\mv@normal\upmath@group{eur}{m}{n}
      \define@mathgroup\mv@bold\upmath@group{eur}{b}{n}
      \edef\UPM{\hexnumber\upmath@group}
      \new@mathgroup\amsa@group
      \define@mathgroup\mv@normal\amsa@group{msa}{m}{n}
      \define@mathgroup\mv@bold\amsa@group{msa}{m}{n}
      \edef\AMSa{\hexnumber\amsa@group}
      \makeatother
      \mathchardef\upi="0\UPM19
      \mathchardef\umu="0\UPM16
      \mathchardef\upartial="0\UPM40
      \mathchardef\leqslant="3\AMSa36
      \mathchardef\geqslant="3\AMSa3E

      \let\leq=\leqslant 
      \let\geq=\geqslant \let\ge=\geqslant
    \fi
  \fi
\fi % End of NFSS release 1

\ifnfsstwo
  \DeclareMathAlphabet{\mathbfit}{OT1}{cmr}{bx}{it}
  \SetMathAlphabet\mathbfit{bold}{OT1}{cmr}{bx}{it}
  \DeclareMathAlphabet{\mathbfss}{OT1}{cmss}{bx}{n}
  \SetMathAlphabet\mathbfss{bold}{OT1}{cmss}{bx}{n}
  \ifAMStwofonts
    \ifCUPmtlplainloaded \else
      \DeclareSymbolFont{UPM}{U}{eur}{m}{n}
      \SetSymbolFont{UPM}{bold}{U}{eur}{b}{n}
      \DeclareSymbolFont{AMSa}{U}{msa}{m}{n}
      \DeclareMathSymbol{\upi}{0}{UPM}{"19}
      \DeclareMathSymbol{\umu}{0}{UPM}{"16}
      \DeclareMathSymbol{\upartial}{0}{UPM}{"40}
      \DeclareMathSymbol{\leqslant}{3}{AMSa}{"36}
      \DeclareMathSymbol{\geqslant}{3}{AMSa}{"3E}

      \let\leq=\leqslant 
      \let\geq=\geqslant \let\ge=\geqslant
    \fi
  \fi
\fi % End of NFSS release 2

\ifCUPmtlplainloaded \else
  \ifAMStwofonts \else % If no AMS fonts
    \def\upi{\pi}
    \def\umu{\mu}
    \def\upartial{\partial}
  \fi
\fi

\title{Disk-Halo Interaction - I. Three-Dimensional Evolution of the Galactic Disk}
\author[M. A. de Avillez]
  {Miguel A. de Avillez\thanks{Presently at the Department of Astrophysics, American Museum of Natural History, Central Park West at 79th Street, New York, N.Y. 10024-5192, USA.}\\
  Department of Mathematics, University of \'Evora, R. Rom\~ao Ramalho 59, 7000 \'Evora, Portugal\\
mavillez@galaxy.lca.uevora.pt}

\date{Accepted 1999 *** **.
      Received 1999 *** **;
      in original form 1999 January 11}

\pagerange{\pageref{firstpage}--\pageref{lastpage}}
\pubyear{1999}

\begin{document}

\maketitle

\label{firstpage}

\begin{abstract}
The results of a three-dimensional model for disk-halo interaction are presented here. The model considers explicitly the input of energy and mass by isolated and correlated supernovae in the disk. Once disrupted by the explosions, the disk never returns to its initial state. Instead it approaches a state where a thin HI disk is formed in the Galactic plane overlayed by thick HI and HII gas disk with scale heights of 500 pc and of 1 to 1.5 kpc, respectively. The upper parts of the thick HII disk (the diffuse ionized medium) act as a disk-halo interface and its formation and stability are directly correlated to the supernova rate per unit area in the simulated disk. 
\end{abstract}


\section{Introduction}
This is the first of three papers where the dynamics of the Galactic fountain and its effects in the Galaxy are presented. The present paper describes three-dimensional hydrodynamical simulations of the disk gas powered by isolated and clustered supernovae and its interaction with the halo. 

In paper II (Avillez, 1999b) a discussion is conducted on the dynamics of the Galactic fountain in the halo and the formation of HI clouds condensing from the fountain flow. In paper III a discussion of the present day modelling of the disk-halo interaction is carried out. This includes a detailed presentation and comparison between the existing models, their strengths and weaknesses and how their predictions compare with observations. 

The approach adopted here assumes the Galactic fountains are intimately related to the vertical structure of the thick gas disk and to the rate occurrence of supernovae per unit area in the disk. The hot gas released by the explosions breaks through the thick disk, escaping into the halo via the diffuse ionized medium, and forming large scale fountains.

The term ``galactic fountain'' has been used to define a global circulation model where the disk gas rises up into the halo, cools and returns to the disk. The returning cold gas contitutes some of the observed HI clouds in the halo (see Wakker \& van Woerden, 1997). The latest developments on the study of the physics of the HI gas in the Galactic halo are reviewed in Wakker {\it et al.} (1999). 

Several mechanisms have been proposed for the disk-halo-disk circulation, taking into account the vertical structure of the Galactic disk and the distribution of O and B stars in the disk. These comprise two major models: the classical fountain (CF) model (Shapiro \& Field, 1976; Bregman, 1980; Kahn, 1981) and the chimney model (Tomisaka \& Ikeuchi, 1986; Norman \& Ikeuchi, 1989). 

The CF models assumed the fountain gas being originated in type II supernovae randomly distributed in the disk. As a consequence the gas gains enough energy to expand buoyantly into the halo where it cools and condenses into clouds that would rain over the disk. The maximum velocity these clouds would have is 70 km/s (Kahn, 1981), ruling out HVCs (clouds with descending velocities greater than 100 km/s) as a product of a Galactic fountain. These models did not take into account the presence of a thick disk composed of a multiphase medium with different scale heights commonly known as the Lockman (with a scale height $h_{z}\sim 500$ pc) (Lockman, 1984; Lockman {\it et al.}, 1986) and the Reynolds (with a scale height $h_{z}\sim 1-1.5$ kpc) (Reynolds, 1987) layers.  The presence of such a thick gas disk constrains the dynamics of the hot intercloud medium on its way out of the disk.

The detection of holes with sizes varying between a few hundred and a kilo parsec in the emission maps of nearby spiral galaxies (M31 and M33) by Brinks \& Bajaja (1986) and Deul \& Hartog (1990) and in the Milky Way by Heiles (1984) and Koo {\it et al} (1992) suggest that associations of O and B stars exploding in a correlated fashion could be responsible for the formation of superbubbles that blow holes in the disk. As a consequence, they release their hot inner parts into the lower halo through tunnels physically connected to the underlaying OB associations resembling chimneys (Tomisaka \& Ikeuchi (1986), Norman \& Ikeuchi (1989), Mac Low {\it et al.}, 1989). 

The energies involved in such a phenomenon are of the order of $10^{53}$ erg (Heiles, 1991). As the hot gas rises up to a height of 10 kpc, it cools and returns to the Galactic disk, forming a fountain. The returning clouds would then acquire high velocities. Furthermore, chimneys reduce the volume filling factor of the hot gas in the disk reducing it from 0.7 predicted by the three-phase model of McKee \& Ostriker (1977) to some 0.2 as predicted by Norman \& Ikeuchi (1989). 

Chimneys have been regarded as the most efficient way to expand the hot disk gas into the halo and as indirectly being reponsible for the formation of HVCs (Kuntz \& Danly, 1996; see reviews by Wakker \& van Woerden 1997). The major drawback is the number of OB associations required to produce chimneys in such a way as to guarantee a continuous injection of hot gas into the halo.

Rosen {\it et al.} (1993) and Rosen \& Bregman (1995) developed an integrated model where two co-spatial fluids representing the stars and gas in the interstellar medium were used. These models reproduced the presence of a multiphase media with cool, warm and hot intermixed phases having mean scale heights compatible with those observed in the Galaxy. However, the structure and properties of the ISM changed according to the overall energy injection rate. The best range of supernova rates that allow the reproduction of the ISM stratification are closed to the observed in the Galaxy. The resulting volume filling factors ($50\%$ for the hot medium, $25-30\%$ for the warm gas and $20\%$ for the cold gas (Rosen {\it et al.}, 1999)) are slightly different from those expected for the Galactic disk (see Spitzer, 1990; Ferri\`ere, 1995). 

Substantial progress has been made in modelling the disk-halo interaction and also the Galactic fountain since it was first suggested by Shapiro \& Field in 1976. The models described above give a two-dimensional description of the dynamics of the disk and halo gas, and simulate the vertical distribution of IVCs and HVCs in the halo. Simplified conditions for the stars and ISM have been taken into account. 

The Galactic fountain models assumed that supernovae were randomly distributed in the disk and occurred with a rate of 3 per century, corresponding to a massflux of hot ($T\sim 10^{6}$ K) material into the halo of the order of $10^{-19}$ g cm s$^{-1}$, whereas global models such as those of Rosen \& Bregman (1995) showed that it is very difficult to reproduce, in the same simulation, both the structure of the ISM near the disk and, at the same time, the presence of HI gas with high and intermediate velocities in the halo without increasing the volume filling factor of the hot gas in the disk. 

Both the CF and the global models of the disk-halo interaction show that large scale outflows may be responsible for the formation of some of the observed IVCs and HVCs, providing the hot intercloud medium has a temperature of T$_{\circ}=10^{6}$ K and the density $\rho=1.67\times 10^{-24}$ g cm$^{-3}$. This corresponds to mass influx of HVCs onto the disk of $2-4 \msolar$ yr$^{-1}$. A result in accordance with the estimates carried out by Wakker (1990). The chimney models neglected the presence of a dynamical thick disk which would provide constraints to the structure of the chimneys. The formation of the chimneys was carried out in a medium where no other phenomenon was present. 

Furthermore, the two-dimensional calculations impose natural limitations on the evolution of the gas and clouds. The absence of a third dimension constrains the motion to a vertical plane perpendicular to the Galactic disk. If three-dimensional evolution is considered, the overall structure of the flow may suffer modifications that lead to the generation of phenomena that may not be identified in two-dimensions. 

The principal objective of the research reported here has been to develop a three-dimensional model that can account for the collective effects of type Ib, Ic and II supernovae on the structure of the interstellar medium in the galactic disk and halo. This model should, accordingly, account for the formation of the major features already observed in the Galaxy; these include the Lockman and Reynolds layers, large scale outflows, chimneys, and HI clouds, features that all contribute to the overall disk-halo-disk cycle known as the Galactic fountain.

Section 2 deals with numerical modelling model where the model of the Galactic disk, rate of supernovae , boundary and initial conditions and numerical scheme are presented. Section 3 presents the evolution of the simulations, comprising the global evolution, the description of the simulations as they reach the steady state and the distribution of cold gas in the simulated disk. In section 4, a discussion on the simulations and the major predictions is carried out, and a comparison with existing observations is made. Finally, section 5 presents the conclusions and future prospects.
\begin{table*}
\label{tab1}
\begin{center}
\begin{tabular}{c|c|c|c|c}
\hline
Volume & Distribution & $\sigma_{Ib+Ic}$ & $\sigma_{II}$ & $\large \sigma_{Ib+Ic+II}$\\
 & & \small ($10^{-3}$ yr$^{-1}$) & \small ($10^{-3}$ yr$^{-1}$) & \small ($10^{-3}$ yr$^{-1}$)\\
\hline
\hline
\multicolumn{1}{c|}{Galaxy} & OB Assoc. & 1.2 & 7.2 & 8.4\\ 
\cline{2-5}
			    & Isolated 	& 0.8 & 4.8 & 5.6 \\
\hline
\hline
\multicolumn{1}{c|}{Sim. Disk} & OB Assoc. & 0.002 & 0.01 & 0.012\\ 
\cline{2-5}
& Isolated & 0.001& 0.007 & 0.008\\
\hline
\end{tabular}
\caption{Rates of occurrence of supernovae types Ib+Ic and II in the Galactic disk having a volume $V_{G}$ and in the simulated stellar disk with volume $V$. The rates account for isolated as well as clustered supernovae.}
\end{center}
\end{table*}
%%%%%%%%%%%%%%%%%%%%%%%%%%%%%%%%%%% SECTION 2 %%%%%%%%%%%%%%%%%%
\section{Numerical Modelling}
\subsection{Model of the Galaxy}
The study of the evolution of the Galactic disk rests in the realization that the Milky Way has a thin and a thick disk of gas in addition to a stellar disk. The thin gas disk has a characteristic thickness comparable to that of the stellar disk of Population I stars. The thick gas disk is composed of warm neutral and ionized gases with different scale heights - 500 pc (Lockman {\it et al.}, 1986) and 950 pc (Reynolds, 1987), respectively. 

The stellar disk has a half thickness of 100 pc and the vertical mass distribution, $\rho_{\star}$, inferred from the star kinematics, given by (Avillez {\it et al.}, 1998),
\begin{equation}
\label{eq1}
\rho_{\star}=\rho_{\star,\circ} sech ^{2}\left[\left(2\pi G \beta_{\star}\rho_{\star}\right)^{1/2}z\right]
\end{equation}
where $z$ varies between $-100$ pc and $100$ pc, $\rho_{\star,\circ}=3.0\times 10^{-24}$ g cm$^{-3}$ is the mass density contributed by Population I stars near the Galactic plane (Allen, 1991) and the constant $\beta_{\star}=1.9\times 10^{-13}$ cm$^{-2}$ s$^{2}$. This mass distribution generates a local gravitational potential, $\Phi$, of the form
\begin{equation}
\label{eq2}
\Phi=-\frac{2}{\beta_{\star}}\ln \cosh\left[\left(2\pi G\rho_{\star}\beta_{\star}\right)^{1/2}z\right].
\end{equation}

The stellar disk is populated by supernovae types Ib, Ic and II, whose major fraction occurs along the spiral arms of the Galaxy close to or in HII regions (Porter \& Filippenko, 1987). Supernovae types Ib and Ic have progenitors with masses $M\geq 15\msolar$, whereas Type II SNe originate from early B-type precursors with $9 \msolar\leq M\leq 15 \msolar$. 

The rates of occurrence of supernovae types Ib+Ic and II in the Galaxy are $2\times 10^{-3}$ yr$^{-1}$ and $1.2\times 10^{-2}$ yr$^{-1}$ respectively (Cappellaro {\it et al.}, 1997). Similar rates have been found by Evans {\it et al.} (1989) in a survey of 748 Shapley-Ames galaxies. The total rate of these supernovae in the Galaxy is $1.4\times 10^{-2}$ yr$^{-1}$. Sixty percent of these supernovae occur in OB associations, whereas the remaining $40\%$ are isolated events (Cowie {\it et al.}, 1979). 

\subsection{Basic Equations and Numerical Methods} 
The evolution of the disk gas is described by the equations of conservation of mass, momentum and energy:
\begin{equation}
\label{eq3}
\frac{\partial \rho}{\partial t}+\nabla\left(\rho {\bf v}\right)=0;
\end{equation}
\begin{equation}
\label{eq4}
\frac{\partial\left( \rho {\bf v}\right)}{\partial t}+{\bf \nabla}\left(\rho {\bf v}{\bf v}\right)=-{\bf \nabla}p-\rho {\bf \nabla}\Phi;
\end{equation}
\begin{equation}
\label{eq5}
\frac{\partial \left(\rho e\right)}{\partial t}+{\bf \nabla}\left(\rho e {\bf v}\right)=-p{\bf \nabla}{\bf v}-n^{2}\Lambda;
\end{equation}
where $\rho$, $p$, $e$, and ${\bf v}$ are the mass density, pressure and specific energy and velocity of the gas, respectively. The set of equations is complete with the equation of state of ideal gases. 

$\Lambda$ is the functional approximation of the cooling functions of Dalgarno \& McCray (1972), Raymond {\it et al.} (1976) and the isochoric curves of Shapiro \& Moore (1976), except in the range of temperatures of $10^{5}-5\times 10^{6}$ K where the simple power law of the temperature (Kahn, 1976)
\begin{equation}
\Lambda=1.3\times 10^{-19} T^{-0.5}\quad\mbox{erg cm$^{3}$ s$^{-1}$}
\end{equation}
is applied. Between $5\times 10^{6}$ and $5\times 10^{7}$ the cooling function has been approximated by $T^{-0.333}$ (Dorfi, 1997). In order to avoid any cooling below zero temperature, $\Lambda$ is zero below 200 K.

The equations of evolution are solved by means of a three-dimensional hydrodynamical scheme using the piecewise parabolic method (Collela \& Woodward, 1984) and the adapted mesh refinement algorithm of Berger \& Collela (1989). Three levels of refinement are used. Each is created with a refinement factor of two. The maximum resolution obtained with this method is 1.25 pc.

\subsection{Computational Domain and boundary Conditions}
The simulations were carried out using a cartesian grid centred on the Galactic plane with an area of 1 kpc$^{2}$ and extending from -4 kpc to 4 kpc. Grid resolutions of 5 pc and 10 pc were used for $-270\leq z\leq 270$ pc and for $z\leq -250$ pc and $z\geq +250$ pc, respectively. The cells located at $-270\leq z\leq -250$ pc and $250\leq z\leq 270$ pc of the coarser and finer grids overlap to ensure continuity between the two grids. The solution in the finer grid, at ranges $-270\leq z\leq -250$ pc and $250\leq z\leq 270$ pc, results from a conservative interpolation from the data in the coarser cells to the fine grid cells as prescribed by Berger \& Collela (1989). 

The boundary conditions along the vertical axis are periodic; outflow boundary conditions are used in the upper and lower parts of the grid parallel to the Galactic plane.
\begin{figure*}
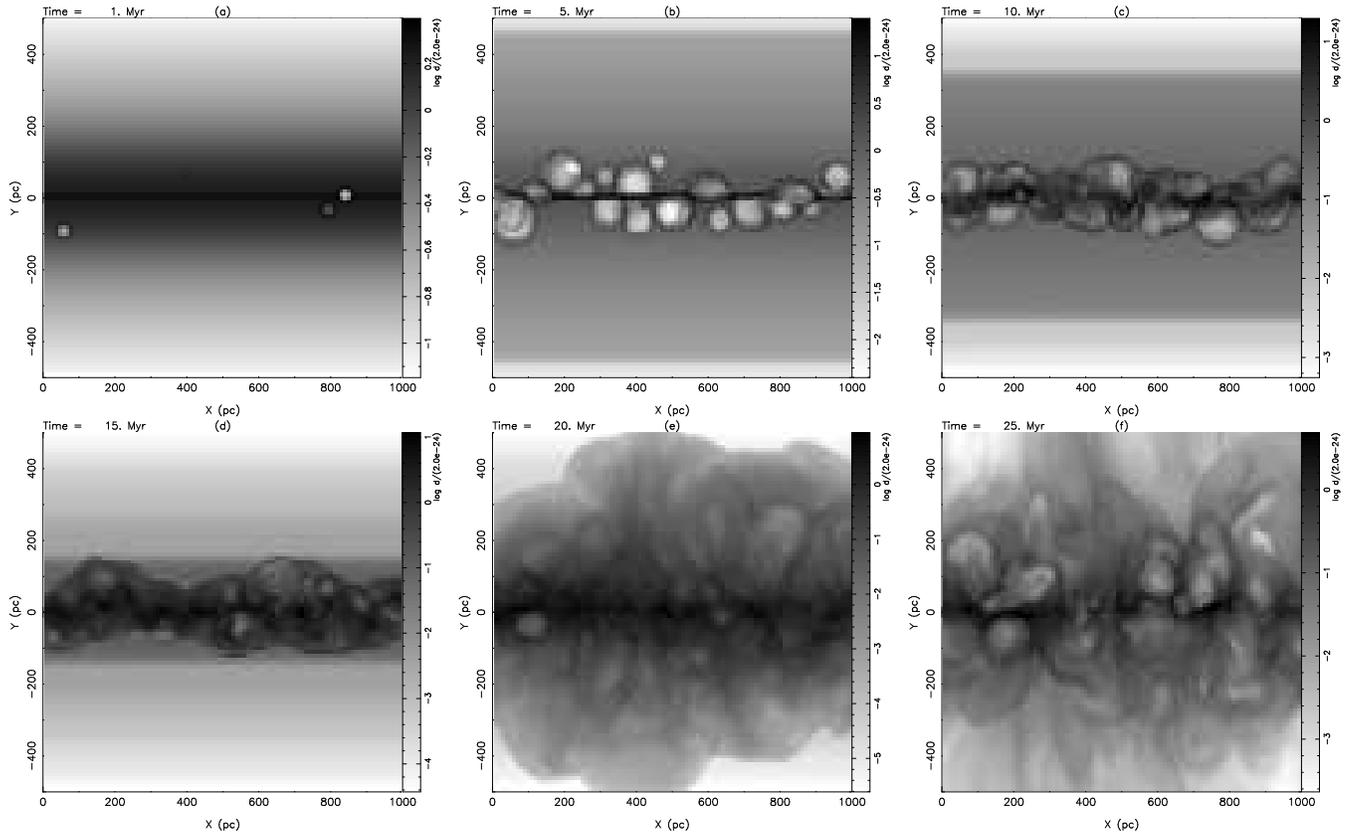

\centering{\hspace*{-0.1cm}}
\psfig{file=mavillez_fig1-abc.ps,angle=-90,width=\hsize,clip=}
\psfig{file=mavillez_fig1-def.ps,angle=-90,width=\hsize,clip=}
\caption{Grey-scale maps of the density distribution during the first 25 Myr for $y=500$ pc at: (a) 1 Myr, (b) 5 Myr, (c) 10 Myr, (d) 15 Myr, (e) 20 Myr and (d) 25 Myr after start of calculations. Note the collapsing of the gas during the first 15 Myr, forming a layer containing the major fraction of the initial mass. As the cold gas descends towards the plane it leaves behind a low density medium (light grey in images (b) and (c)).} 
\label{mavillez_fig1}
\end{figure*}
\subsection{Initial Conditions}
The initial configuration of the system in the computational domain considers the vertical distribution of the disk and halo gases in accordance with the Dickey \& Lockman (1990) profile, and reproduces the distribution of the thin and thick disk gas in the solar neighborhood. The gas is initially in hydrostatic equilibrium with the gravitational field, thus requiring an initial gas temperature as defined by the density distribution.

\subsection{Supernovae in the Simulated Disk}
Throughout the simulations the presence of the stellar disk is required with a thickness of 200 pc and an area of 1 kpc$^{2}$, corresponding to a volume of $V=2\times 10^{8}$ pc$^{3}$. Isolated and clustered supernovae of types Ib, Ic and II occur in the volume $V$ at rates obtained from the values discussed in \S2.1 and given by   
\begin{equation}
\sigma\frac{V}{V_{G}}=1.42\times 10^{-3}\sigma \mbox{\hspace*{0.2cm} yr$^{-1}$} \end{equation}
where $V_{G}=1.4\times 10^{11}$ pc$^{3}$ is the volume of the Galactic disk with a radius of 15 kpc and a thickness of 200 pc, and $\sigma$ is the rate of supernovae observed in the volume $V_{G}$. Table 1 presents the rates of supernovae in the Galaxy as well as in the simulated disk. In the simulated stellar disk, isolated and clustered supernovae form at time intervals of $1.26\times 10^{5}$ yr and $8.4\times 10^{4}$ yr respectively.

The time interval between two sucessive superbubbles in the volume $V$ varies between $1.9\times 10^{6}$ yr and $5.23\times 10^{6}$ yr (Ferri\`ere, 1995), but following Norman and Ikeuchi (1989), and adopting an average value of $N_{SN}=30$ supernovae per superbubble, the time interval of formation of superbubbles in the volume $V$ is
\begin{equation}
\frac{N_{SN}}{\sigma_{OB}}=2.5\times 10^{6}\;\mbox{yr},
\end{equation}
the value used in this study.

Each supernova is set up in the beginning of phase II, with a thermal energy content of (Kahn, 1975)
\begin{equation}
E_{Therm}=0.36\rho_{\circ}a
\end{equation}
and a kinetic energy content of
\begin{equation}
E_{Kin}=0.138\rho_{\circ}a,
\end{equation}
where $a=2E/\rho_{\circ}$, $\rho_{\circ}$ is the local density of the medium where the supernova occurs, and $E$ is the energy of the explosion. 

The generation of supernovae in the stellar disk is carried out in a semi-random way, where the coordinates of the supernova are determined from three new random numbers converted to the dimensions of the stellar disk. After the determination of the random numbers, and therefore, the location of the new supernova in the grid, several tests are carried out in order to determine the type of supernova whether it occurs isolated in space or within an OB association, or even to determine if the new location must be dismissed. 

The location of a new supernova in the grid is based on the following constraints: (a) vertical density distribution of population I stars given by equation (\ref{eq1}), (b) type of new supernova, (c) rates of occurrence of supernovae types Ib+Ic and II and (d) rate of formation of superbubbles. 

When a supernova is due to appear in a specific location, a supernova generator checks for the occurrence of a previous supernova in that location. If this test is negative, then a new supernova is set up, providing its generation is compatible with the time interval between the previous supernova of the same type. If the test is positive, then the time delay between the old supernova and the new one is checked. Now the decision is based on the comparison of the this time delay and that of a new supernova within a superbubble. If the delay is smaller than the time delay within the superbubble, then no supernova is generated and the generator chooses a new position from a new set of random numbers, otherwise the supernova is set up.

When the type of the new supernova is known, the algorithm determines the radius of the supernova in the beginning of phase II. This is acomplished by equating the amount of mass of the progenitor star released during its explosion and the mass of the ISM engulfed by the blast wave. The mass of the progenitor star is a ramdom value determined from the interval of masses characteristic to the progenitors of that type of supernova, e.g. if the next supernova to occur is a type Ib, then the amount of mass that is released during the explosion is found within the interval $[15, 30]$ $\msolar$.

%%%%%%%%%%%%%%%%%%%  SECTION 3 - Evolution of the Simulations  %%%%%%%%%%%%%%
\section{Evolution of the Simulations}
The simulations were carried out for a period of 1 Gyr with supernovae being generated at time 0. The simulations start from a state of hydrostatic equilibrium breaking up during the first stages of the simulation. The system evolves into a statistical steady equilibrium where the overall structure of the ISM seems similar on the global scale. 
\subsection{Global Evolution}
During the first 50 Myr of simulated time there is an imbalance between cooling and heating of the gas in the disk. Initially there is an excess of radiative cooling over heating because of the small number of supernovae during this period. 

The gas originally located in the lower halo starts cooling and moves ballistically towards the midplane, colliding there with gas falling from the opposite side of the plane (Figure \ref{mavillez_fig1}). After 15 Myr of evolution, the major fraction of the initial mass is confined to a slab having a characteristic thickness of 100-150 pc (Figure \ref{mavillez_fig1} (d)). A thin wiggly disk of cold gas forms at the midplane and has a thickness of few tens of parsec. As the supernovae occur, they warm up the gas in the slab, gaining enough energy to overcome the gravitational pull of the disk and, therefore, expanding upwards (Figure \ref{mavillez_fig1} (e) and (f)) redistributing matter and energy in the computational domain.

\begin{figure}
\centering
\mbox{\epsfxsize=3.5in\epsfysize=4in\epsfbox{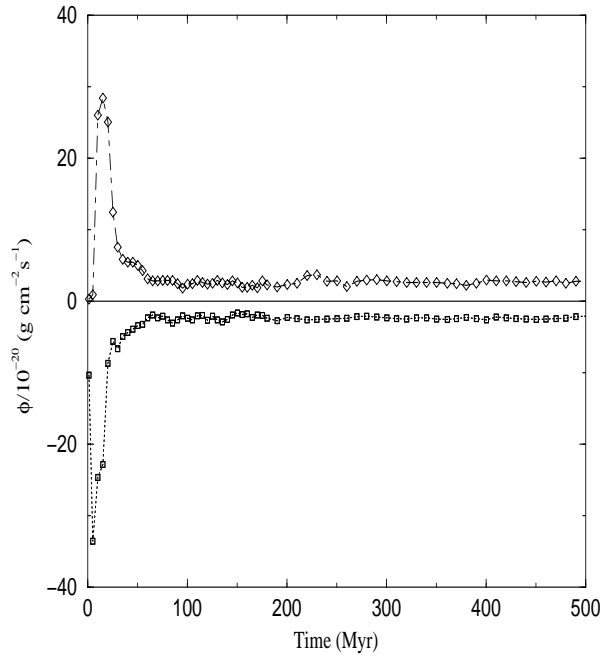}}
\caption{Total massflux $\phi=\rho v_{z}$ of the ascending ($v_{z}>0$, shown as diamonds) and descending ($v_{z}<0$, shown as squares) flows measured at $z=140$ pc and averaged over the entire area of the Galactic disk during the first 500 Myr of evolution. A balance between the ascending and descending flows sets in after 75 Myr.}
\label{mavillez_fig2}
\end{figure}

After the over-expansion, a dynamic equilibrium is set up between upward and downward flowing gas (Figure \ref{mavillez_fig2}). During the first 10-20 Myr, the descending gas has a maximum value of $3.4\times 10^{-19}$ g cm$^{-2}$ s$^{-1}$ (that is $4.8\times 10^{-2}$ $\msolar$ kpc$^{-2}$ yr$^{-1}$). The ascending gas massflux peaks at $2.8\times 10^{-19}$ g cm$^{-2}$ s$^{-1}$ ($3.9\times 10^{-2}$ $\msolar$ kpc$^{-2}$ yr$^{-1}$). The expansion of the disk gas diminishes during the next 20 Myr, with the consequent decrease in the total massflux. Between 60 and 75 Myr, a balance between descending and ascending flows sets in. Both the ascending and descending gases have massfluxes of approximately $3\times 10^{-20}$ g cm$^{-2}$ s$^{-1}$ ($4.2\times 10^{-3}$ $\msolar$ kpc$^{-2}$ yr$^{-1}$), which corresponds to a total inflow rate of 2.97 $\msolar$ yr$^{-1}$ in the Galaxy on one side of the Galactic plane.

Such a dynamic equilibrium is observed during the rest of the simulations, indicating that a quasi-equilibrium between the descending and ascending gases dominates the evolution of the disk-halo gas. However, there are some periods when one of the flows dominates over the other. This is related to periodic imbalances between cooling and heating leading to compressions and expansions of the disk gas.

\begin{figure*}
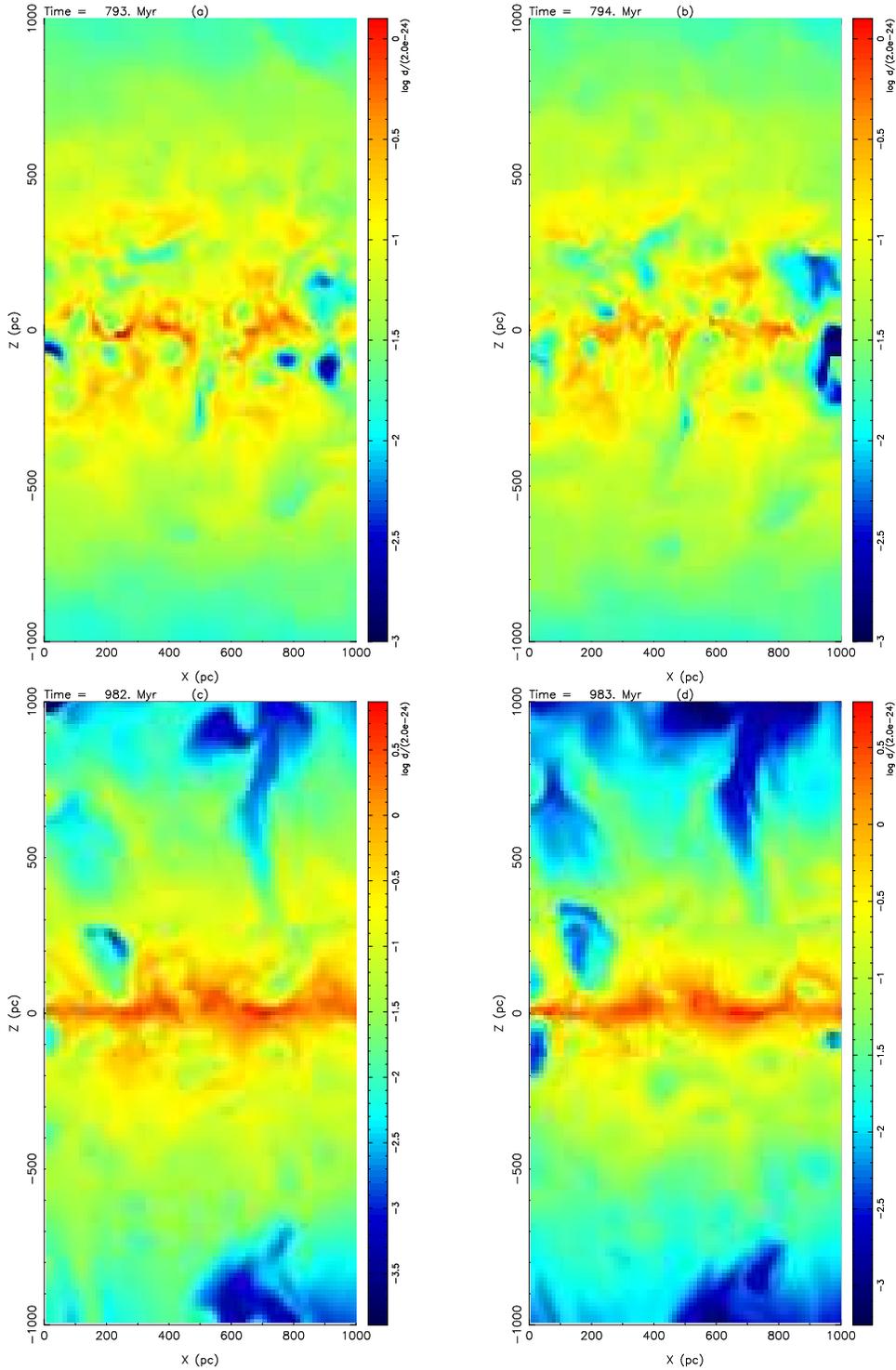

\centering{\hspace*{-0.1cm}}
\psfig{file=mavillez_fig3-ab.ps,angle=-90,width=5in,clip=}
\psfig{file=mavillez_fig3-cd.ps,angle=-90,width=5in,clip=}
\caption{Color maps showing the structure of the Galactic disk measured at two locations in the disk: $y=480$ ((a) and (b)) and $y=800$ pc ((c) and (d)). The times at which these images refer are: (a) 793 Myr, (b) 794 Myr, (c) 982 Myr and (d) 983 Myr. The color scale varies from dark blue (low density gas) to red (high density gas). Neutral gas is represented by a range of colors varying between yellow and red, whereas ionized gas is shown in blue scale. The images show that globally the disk has similar structures, although locally there are variations. Note the global expansion of warm (light blue) and hot (dark blue) ionized gas carving the thick disk and the constant presence of the wiggly cold disk (red) at $z=0$ pc. Local outbursts such as mini-chimneys ((a) and (b), located at $x=500$, $z=-100$ pc) and chimneys ((c) and (d)) are also observed.}
\label{mavillez_fig4}
\end{figure*}

As the balance between descending and ascending flows is reached, the disk evolves into a statistical steady equilibrium where, in general, the disk gas is composed of a multiphase medium extending into the halo, forming a frothy region laced by sheets and crossed by HI clouds and well collimated chimneys as can be seen in Figure \ref{mavillez_fig4}. The figure presents the vertical distribution of the density up to 1 kpc on either side of the plane.

The disk gas shows a stratified distribution where neutral and ionized phases co-exist having different scale-heights. The neutral phase is composed by cold ($T\leq 10^{3}$ K) and warm ($10^{3}<T< 10^{4}$ K) gas whereas the ionized phase is composed by a warmer gas with $10^{4}\leq T< 10^{5}$ K (hereafter identified as warm ionized gas) and hot gas ( $T\geq 10^{5}$ K).

\begin{figure*}
\centering
\psfig{file=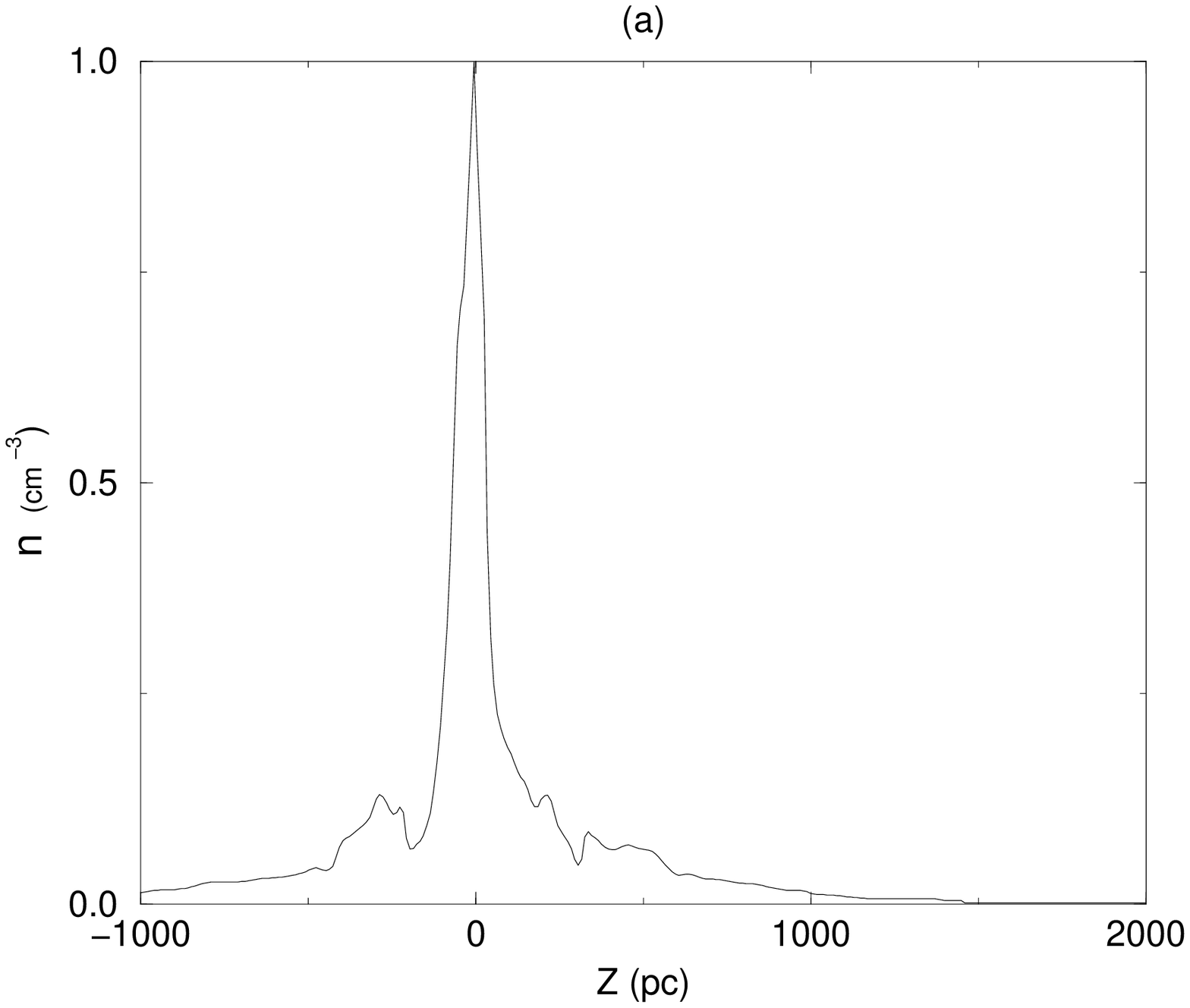,angle=0,width=3.2in,height=2.5in,clip=}
\psfig{file=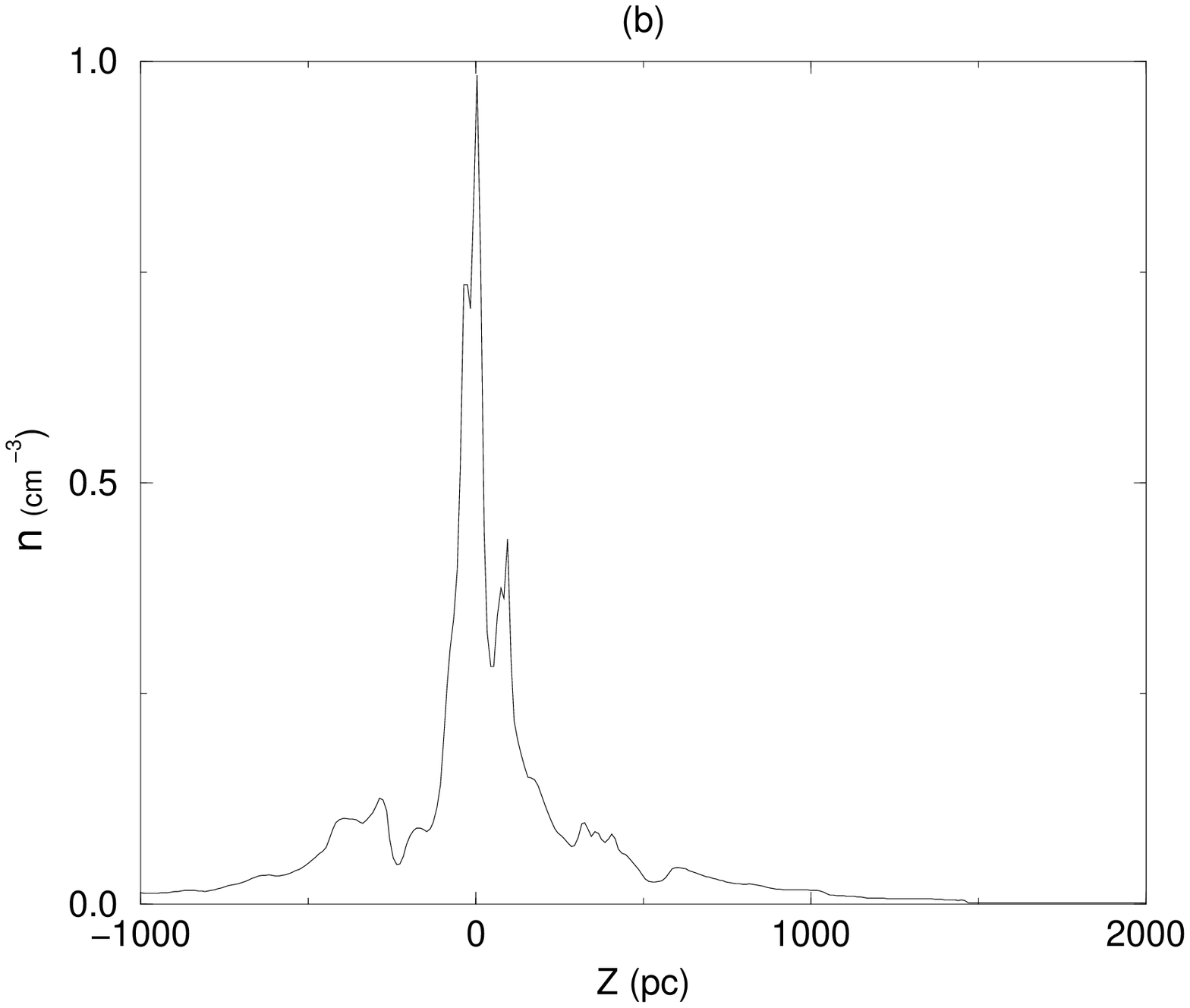,angle=0,width=3.2in,height=2.5in,clip=}
\caption{Density profile of the disk and halo gas measured along two line of sights passing through the plane at: (a) $x=300,\, y=250$ pc and (b) $x=750,\,y=750$ pc. Note the presence of a thin disk with a thickness of 100 pc centred at $z=0$. The density profile behaves like a gaussian distribution up to 200 pc on either side of the plane followed by an exponential up to 1 kpc. Above this height the density decreases steeply showing a separation between the warm ionized medium and the halo gas at 1.5 kpc.}
\label{mavillez_fig5}
\end{figure*}

The neutral component is mainly distributed in the layer $\left|z\right|\leq 500$ pc, forming a thick HI disk and having an average temperature of $10^{4}$ K. Cold gas is present, predominantly in a thin but irregular layer (thin HI disk) around the Galactic plane, having a characteristic thickness of tens of parsec (Figures \ref{mavillez_fig4} and \ref{mavillez_fig5}), whereas the hot gas is mainly distributed above 1.5 kpc. 

\begin{figure*}
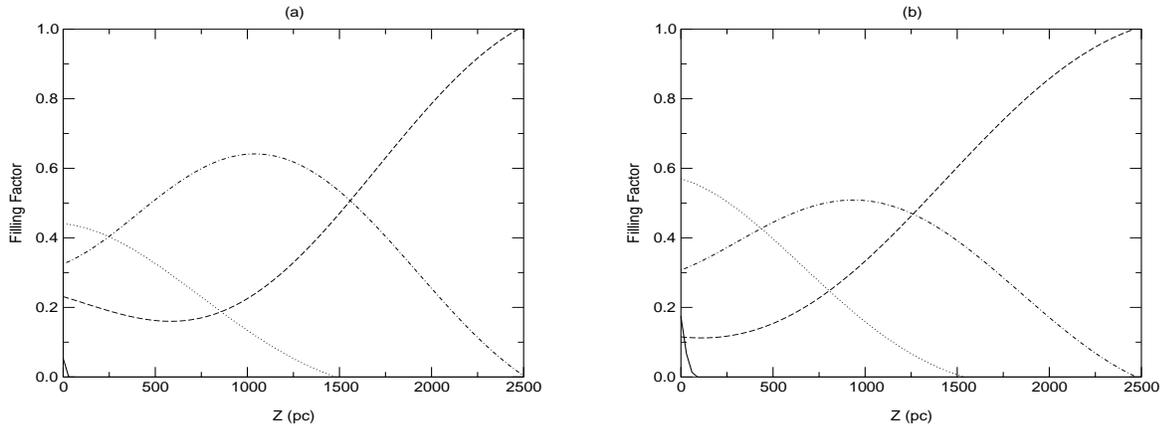

\centering
\psfig{file=mavillez_fig5-a.eps,angle=0,width=2.8in, height=2.2in,clip=}\hspace*{1cm}
\psfig{file=mavillez_fig5-b.eps,angle=0,width=2.8in,height=2.2in,clip=}
\caption{Best fits for distribution in $z$ of the volume filling factors of cold (solid line located at the lower left edge of the plots), warm neutral (dotted line), warm ionized (dot-dashed line) and hot (long dashed line) gas. The fits were calculated from a series of 50 profiles, measured with a time interval of 1 Myr, and taken at: (a) $450-500$ Myr and (b) $950-1000$ Myr. The variation of the volume filling factors with $z$ are compatible with a stratified distribution of the disk and halo gases. The neutral gas dominates for $z\leq 500$ pc, the warm ionized gas is mainly found in the layer located between 500 pc and 1.5 kpc and the hot gas dominates for $z> 1.5$ kpc. The height at which the hot gas becomes dominant is identified as the disk-halo interface.} 
\label{mavillez_fig3}
\end{figure*}

Gas with temperatures varying between $10^{4}$ and $10^{5}$ K fill up to $\sim 80\%$ of the disk volume, whereas gas with temperatures smaller than $10^{4}$ K, fills on average (Figure \ref{mavillez_fig3}), $50\%$ of the disk volume. The $T\leq 10^{4}$ gas is mainly located in a layer with half thickness of 500 pc ($|z|\leq 500$ pc) (see also Figure \ref{mavillez_fig4}), although some of this gas extends to 1.5 kpc, but with a very small volume filling factor. The warm ionized gas dominates the region between 500 pc and 1.5 kpc, although its presence is detected up to 2.5 kpc. Gas with temperatures larger than $10^{5}$ K fill on average $15-23\%$ of the disk volume below 1 kpc. With the increase of $z$, the volume occupied by the hot gas increases, becoming the dominant medium at $z\ge 1.5$ kpc (Figure \ref{mavillez_fig3}). 
The density distribution of the disk gas decreases with $z$. At $\left|z\right|\leq 200$ pc it can be approximated by a gaussian followed by an exponential distribution (Figure \ref{mavillez_fig5}). The gaussian is overshot by the exponential distribution with a smooth decrease followed by a steep increase in the density profile (Figure \ref{mavillez_fig5}) on either side of the plane. This suggests the presence of an interface between the thin disk and the thick HI disk. Such an interface is fed by the hot gas flowing upwards from the thin disk in large scale outflows ($l\sim 200$ pc or more) interacting with the cooler thick disk gas overlaying it. Such a configuration becomes Rayleigh-Taylor unstable, and therefore the upward flow acquires a finger-like structure during its expansion through the thick disk. 

This density distribution gives the appearance of a stratified distribution which is compatible to the volume filling factors of the different phases of the ISM at different $z$ (see Figure \ref{mavillez_fig3}).  
\begin{figure}
\centering
\psfig{file=mavillez_fig6-a.ps,angle=-90,width=2.5in,clip=}\\
\psfig{file=mavillez_fig6-b.ps,angle=-90,width=2.5in,clip=}\\
\psfig{file=mavillez_fig6-c.ps,angle=-90,width=2.5in,clip=}\\
\psfig{file=mavillez_fig6-d.ps,angle=-90,width=2.5in,clip=}\\
\psfig{file=mavillez_fig6-e.ps,angle=-90,width=2.5in,clip=}\\
\psfig{file=mavillez_fig6-f.ps,angle=-90,width=2.5in,clip=}\\
\caption{Gray scale maps of the gas density measure at 500 Myr of disk evolution for: (a) y=30, (b) 220, (c) 410, (d) 600, (e) 790 and (f) 980 pc from the edge of the computational domain. A thin HI disk with a variable thickness and a wiggly structure is present in all the images. Its number density varies between 1 and 50 cm$^{-3}$. The largest density values are found in the midplane stretching over large regions with sizes of $50 - 150$ pc. Cold gas is mainly distributed in the thin disk and in sheet like structures found isolated or connected to the underlying thin HI disk, resembling worms.} 
\label{mavillez_fig6}
\end{figure}
\begin{figure}
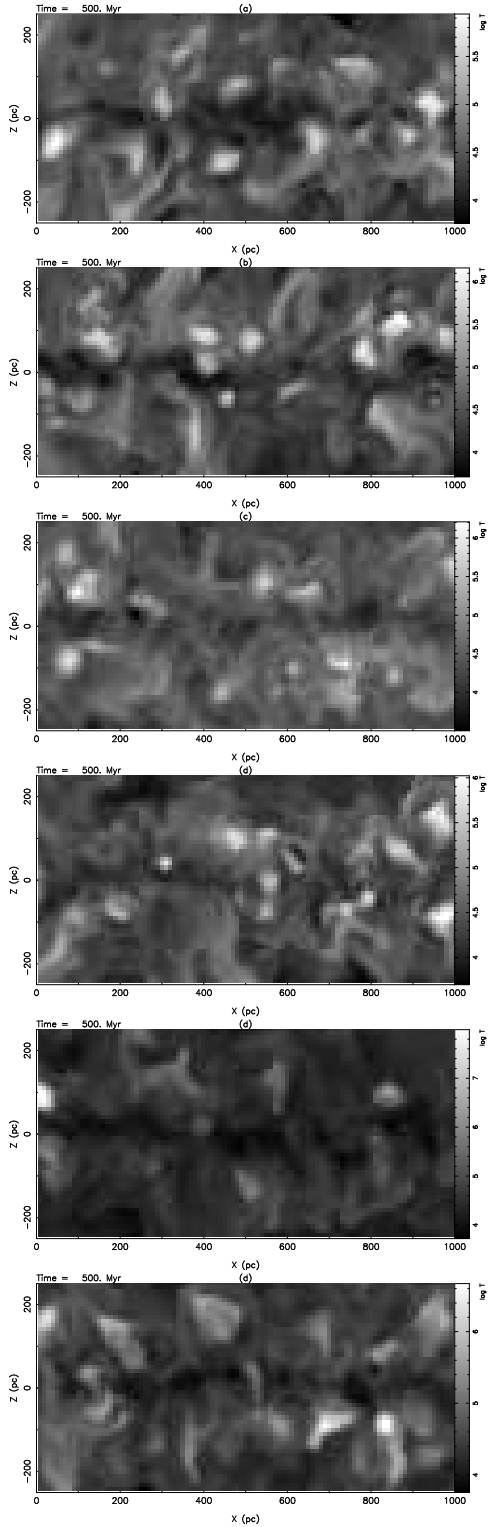

\centering
\psfig{file=mavillez_fig7-a.ps,angle=-90,width=2.5in,clip=}\\
\psfig{file=mavillez_fig7-b.ps,angle=-90,width=2.5in,clip=}\\
\psfig{file=mavillez_fig7-c.ps,angle=-90,width=2.5in,clip=}\\
\psfig{file=mavillez_fig7-d.ps,angle=-90,width=2.5in,clip=}\\
\psfig{file=mavillez_fig7-e.ps,angle=-90,width=2.5in,clip=}\\
\psfig{file=mavillez_fig7-f.ps,angle=-90,width=2.5in,clip=}\\
\caption{Temperature maps of the same regions shown in the last figure. Note the presence of the thin HI disk with temperatures varying between $10^{3}$ and $10^{3.8}$ K. The coldest regions are those with larger densities (compare with previous figure). The major fraction of the disk is filled with gas having temperatures of some $10^{4}$ K. Hot gas is found mainly in regions where supernovae went off and constitutes a network of hot tunnels that may cross the midplane.}
\label{mavillez_fig7}
\end{figure}
\begin{figure}
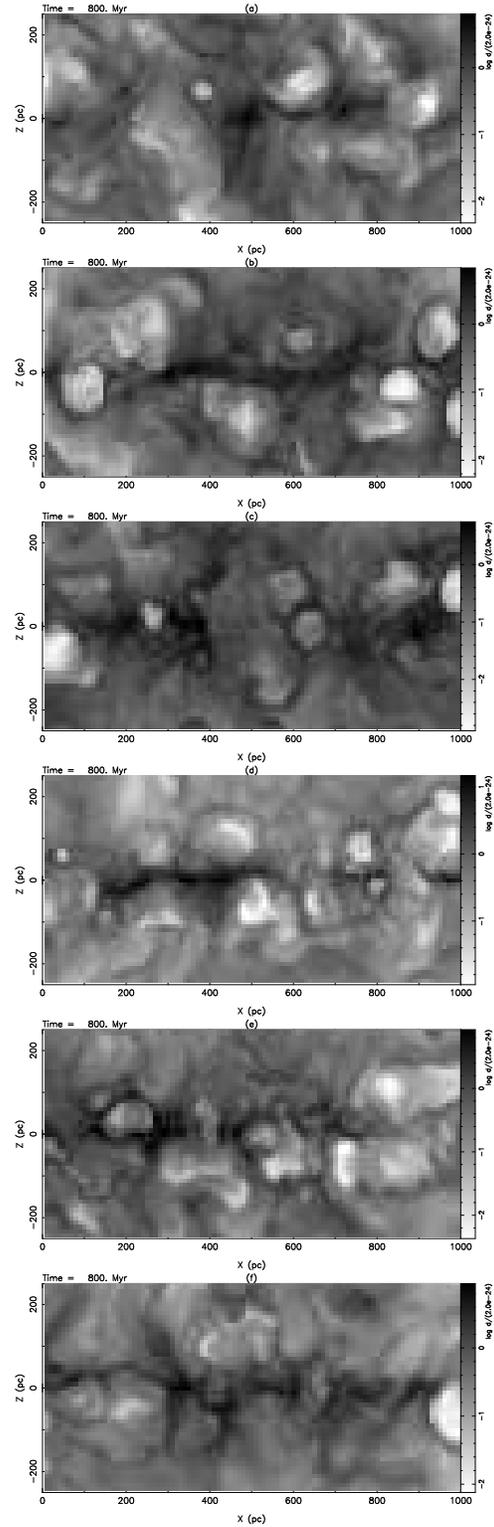

\centering
\psfig{file=mavillez_fig8-a.ps,angle=-90,width=2.5in,clip=}\\
\psfig{file=mavillez_fig8-b.ps,angle=-90,width=2.5in,clip=}\\
\psfig{file=mavillez_fig8-c.ps,angle=-90,width=2.5in,clip=}\\
\psfig{file=mavillez_fig8-d.ps,angle=-90,width=2.5in,clip=}\\
\psfig{file=mavillez_fig8-e.ps,angle=-90,width=2.5in,clip=}\\
\psfig{file=mavillez_fig8-f.ps,angle=-90,width=2.5in,clip=}\\
\caption{Gray scale maps of the gas density measure at 800 Myr of disk evolution for: (a) y=30, (b) 220, (c) 410, (d) 600, (e) 790 and (f) 980 pc from the edge of the computational domain. These images are similar to those shown in Figure \ref{mavillez_fig6}, although locally there are changes due to different events occurring in the disk.} 
\label{mavillez_fig8}
\end{figure}
\begin{figure}
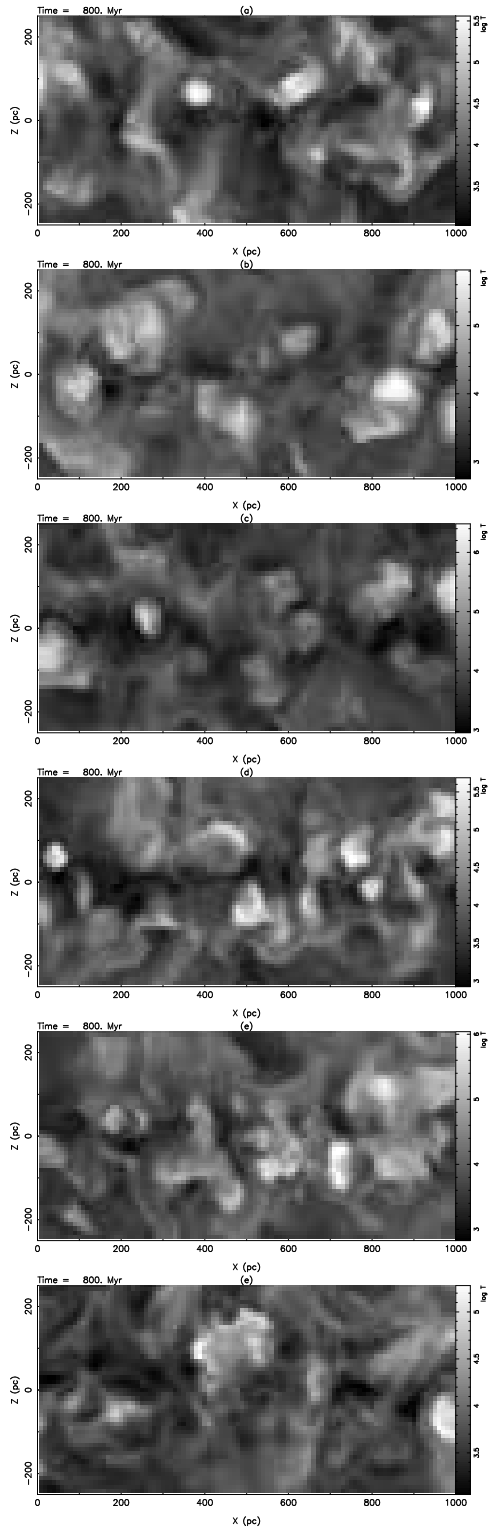

\centering
\psfig{file=mavillez_fig9-a.ps,angle=-90,width=2.5in,clip=}\\
\psfig{file=mavillez_fig9-b.ps,angle=-90,width=2.5in,clip=}\\
\psfig{file=mavillez_fig9-c.ps,angle=-90,width=2.5in,clip=}\\
\psfig{file=mavillez_fig9-d.ps,angle=-90,width=2.5in,clip=}\\
\psfig{file=mavillez_fig9-e.ps,angle=-90,width=2.5in,clip=}\\
\psfig{file=mavillez_fig9-f.ps,angle=-90,width=2.5in,clip=}\\
\caption{Temperature maps of the same regions shown in the last figure. Note the presence of streams of hot gas crossing the midplane, but this is a locallized event resulting from the large number of explosions that occurred in this region in a small period of time.}
\label{mavillez_fig9}
\end{figure}
The ionized component of the disk gas extends upwards up to $z\sim \pm 1.5$~kpc. Its density decreases smoothly up to 1.4-1.5 kpc, where it has a steep decrease. This region acts as an interface between the thick disk and the halo gas. For $z>1.5$ kpc the gas has lower densities and temperatures $T\geq 10^{6}$ K. 
\begin{figure}
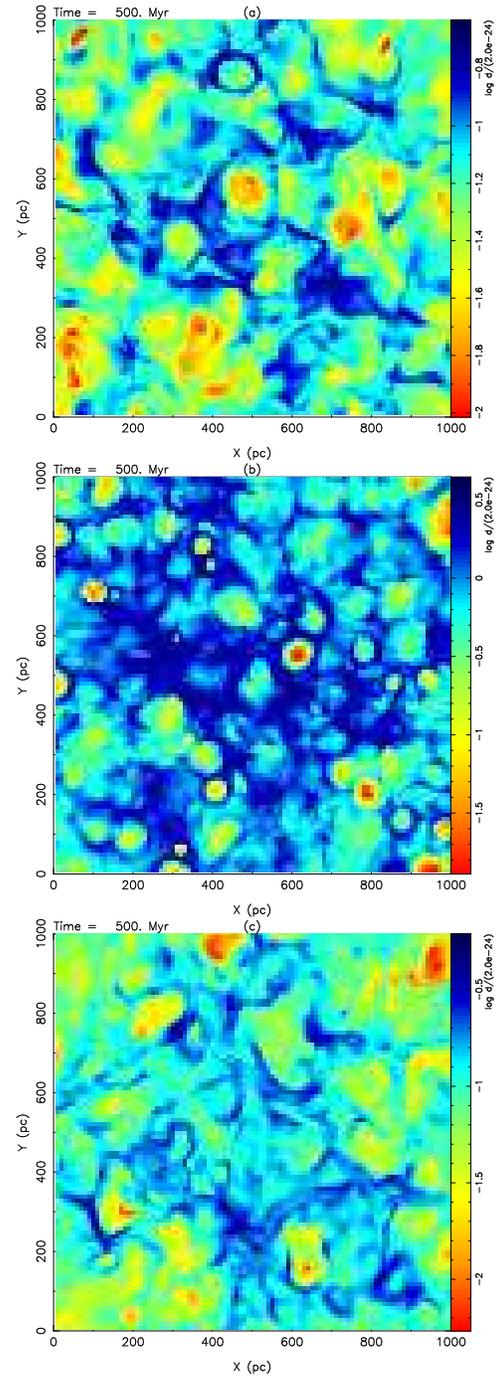

\centering
\psfig{file=mavillez_fig10-a.ps,angle=-90,width=2.5in,clip=}\\
\psfig{file=mavillez_fig10-b.ps,angle=-90,width=2.5in,clip=}\\
\psfig{file=mavillez_fig10-c.ps,angle=-90,width=2.5in,clip=}
\caption{Maps of the density distribution in the disk at 500 Myr and for: (a) $z=-250$, (b) $z=0$ and (c) $z=250$ pc. The HI gas observed at $z=\pm 250$ pc is mainly concentrated in sheet-like structures, connected to others. some of the sheet are the walls of shells and supershells. Some of the cold gas present at $z=0$ is distributed in large regions whose centres have the largest densities - the darkest blue regions in image b).}
\label{mavillez_fig10}
\end{figure}
\begin{figure}
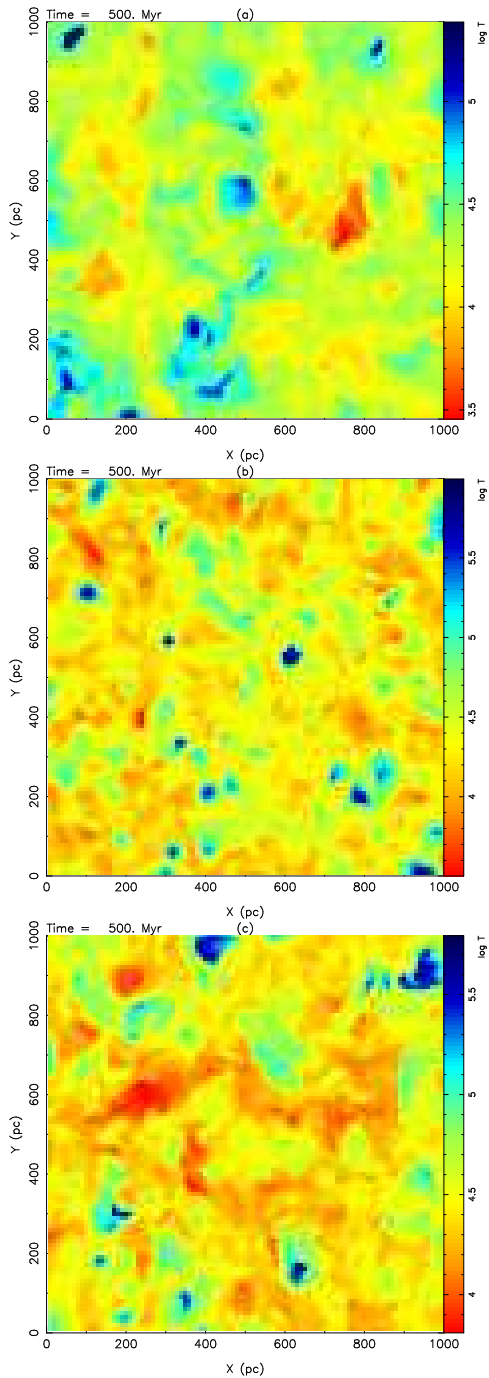

\centering
\psfig{file=mavillez_fig11-a.ps,angle=-90,width=2.5in,clip=}\\
\psfig{file=mavillez_fig11-b.ps,angle=-90,width=2.5in,clip=}\\
\psfig{file=mavillez_fig11-c.ps,angle=-90,width=2.5in,clip=}
\caption{Color maps of the disk temperature at 500 Myr and for: (a) $z=-250$, (b) $z=0$ and $z=250$ pc.}
\label{mavillez_fig11}
\end{figure}

\subsection{Statistical Steady Evolution}
The disk gas shows a steady equilibrium where its structure and dynamics are similar throughout the rest of the simulations. That is, the disk has the same structure at different times and locations, although it experiences changes due to local events such as supernovae, superbubbles, chimneys, collisions with HI clouds, shock waves, etc. In order to stress this point, images showing the time evolution of the disk gas at different locations are presented in Figures \ref{mavillez_fig6}-\ref{mavillez_fig11}. The figures were taken in directions perpendicular and parallel to the midplane at 500 and 800 Myr of evolution time. The $xz-$slices (Figures \ref{mavillez_fig6}-\ref{mavillez_fig9}) were taken at 30, 220, 410, 600, 790 and 980 pc from the edge of the computational domain, whereas the $xy-$slices (Figures \ref{mavillez_fig10}-\ref{mavillez_fig11}) were taken at $z=-250$, $0$ and $250$ pc. The density distributions of the disk gas shown in Figures \ref{mavillez_fig6}, \ref{mavillez_fig8} and \ref{mavillez_fig10} are complemented with the corresponding temperature maps in Figures \ref{mavillez_fig7}, \ref{mavillez_fig9} and \ref{mavillez_fig11}. 

The figures show a large number of features comprising, among others, HI clouds and sheets, bubbles and superbubbles, interacting supernovae forming networks of hot gas, large regions of hot gas crossing the $xy-$slices at $z=-250$ and 250 pc, etc. These features can be summarised as follows:
\begin{itemize}
\item A wiggly thin and a cold disk, having a characteristic thickness varying between a few parsec and tens of parsec (Figures \ref{mavillez_fig6}-\ref{mavillez_fig9}). There are regions where the thin disk may have disappeared as result of a large number of supernovae in the neighborhood as shown in Figure \ref{mavillez_fig8} (a) at $200\leq x\leq 400$ pc. The region is crossed by a stream of hot gas as can be seen in the corresponding temperature map. 
\item Vertical structures connected to the underlying thin disk and appearing to wiggle around like worms. These structures are the debris of broken shells and supershells, that occurr displaced from the midplane. 
\item Sheets of cold gas distributed in the disk that result from the breakup of larger structures (Figures \ref{mavillez_fig6} - \ref{mavillez_fig11}). The sheets have thicknesses of a few pc and widths of several tens of parsec, even hundreds of parsec. The sheets observed at $z=\pm 250$ pc seem to connect to each other, forming a ``network'' of cold gas. In the plane, the cold gas forms large regions of very high density and low temperature (Figure \ref{mavillez_fig10} (b) the darkest blue regions). 
\item Small cloudlets in the vicinity of worms and other sheet-like structures. The smallest cloudlets detected in the images have 1.25 pc size - this corresponds to the maximum resolution used in the disk. The cloudlets result from the breakup of the worms and other clouds which have been swept up by blast waves, involved in cloud-cloud collisions, etc.
\item Supernovae and superbubbles at different phases of their evolution surrounded by shells with variable thicknesses. Supernovae and superbubbles located above the thin gas disk show an elongated appearance as result of the density gradient along the $z-$direction (Figures \ref{mavillez_fig6}-\ref{mavillez_fig8}). 
\item Networks of hot gas connecting supernovae. These can be observed in both the $xz-$slices and $xy-$slices of the temperature distribution of the disk gas. The network of hot gas is not confined to the plane ($z=0$), but spreads in the volume of the disk between $z=-250$ and $250$ pc.
\item Reservoirs of hot gas (large regions of hot gas) moving away from the plane (Figures \ref{mavillez_fig7}, \ref{mavillez_fig8}  and  \ref{mavillez_fig11}). 
\end{itemize}
In general the disk gas is pervaded by isolated supernovae and superbubbles (large supernova structures generated by sucessive supernovae). Supernovae change the local structure of the interstellar medium, but are unable to change the global structure. Individual and clustered supernovae dominate the local environment, depending on their spatial location. Isolated supernovae change the structure of the inner parts of the disk whereas supernovae in OB associations dominate the upper regions of the disk. 

As the blast waves from supernovae expand in the disk, they sweep up clouds and sheets, triggering their disruption into smaller structures. At $z=\pm 250$ pc the presence of sheets results from: (a) cold gas descending from above and acquiring sheet-like forms as a result of Rayleigh-Taylor instabilities that occurr in larger clouds and (b) the breaking up of shells and supershells that expand upwards and are displaced from the Galactic plane.

Young supernovae expand into a highly unstructured medium interacting with other supernovae or cold gas in the form of sheets or clouds (Figures \ref{mavillez_fig6}-\ref{mavillez_fig11}). During their expansion, the remnants are preceeded by shock waves engulfing the shells of other supernovae and triggering their disruption. If the average shock speed in the shell is $v_{s}\sim 1\times 10^{4}$ km s$^{-1}$, then the transmitted shock takes a time 
\begin{equation}
t_{cross}\sim \frac{\Delta R}{v_{s}}
\end{equation} 
to cross the old shell, $\Delta R$ being the thickness of the shell normally 5 to 10 pc thick. Thus, the blast wave crosses the shell in $t_{cross}\leq 10^{4}$ yr, generating a large pressure gradient which is enough to disrupt the old shell. As a consequence, the hot gas flows freely between the remnants forming a network of channels of low density gas. 
\begin{figure*}
\centering
\psfig{file=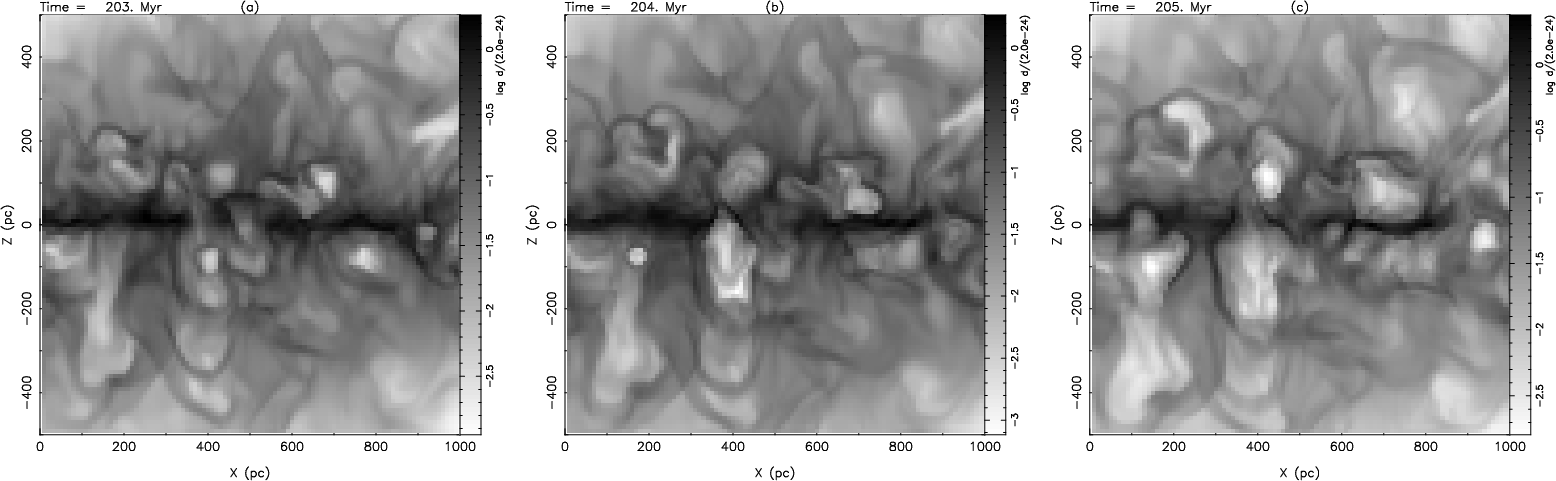,angle=-90,width=\hsize,clip=}
\caption{Evolution of a chimney formed in an OB association located at $(x=400, y=30, 50 \leq z\leq 150)$ pc. The images were taken at: (a) 203 Myr, (b) 204 Myr and (c) 205 Myr of simulated time. This figure is particularly rich in details regarding to the time evolution of the superbubble. After the superbubble has been formed sucessive explosions displaced the upper parts of its shell as can be seen in figure (a) located at $z=\sim -200$ and $-350$ pc. The shell caps have been separated from the shell and conserve their forms. Note the increase in thickness of the left wall of the bubble as sucessive explosions occur. A large plume of hot gas rises near the superbubble at $0\leq x\leq 200$. This structure expands as result of the high energy content in the reservoir of hot gas lying underneath. As the gas flows upwards it acquires a finger-like structure, whereas the chimney conserves its collimated tube-like form.}
\label{mavillez_fig16}
\centering
\psfig{file=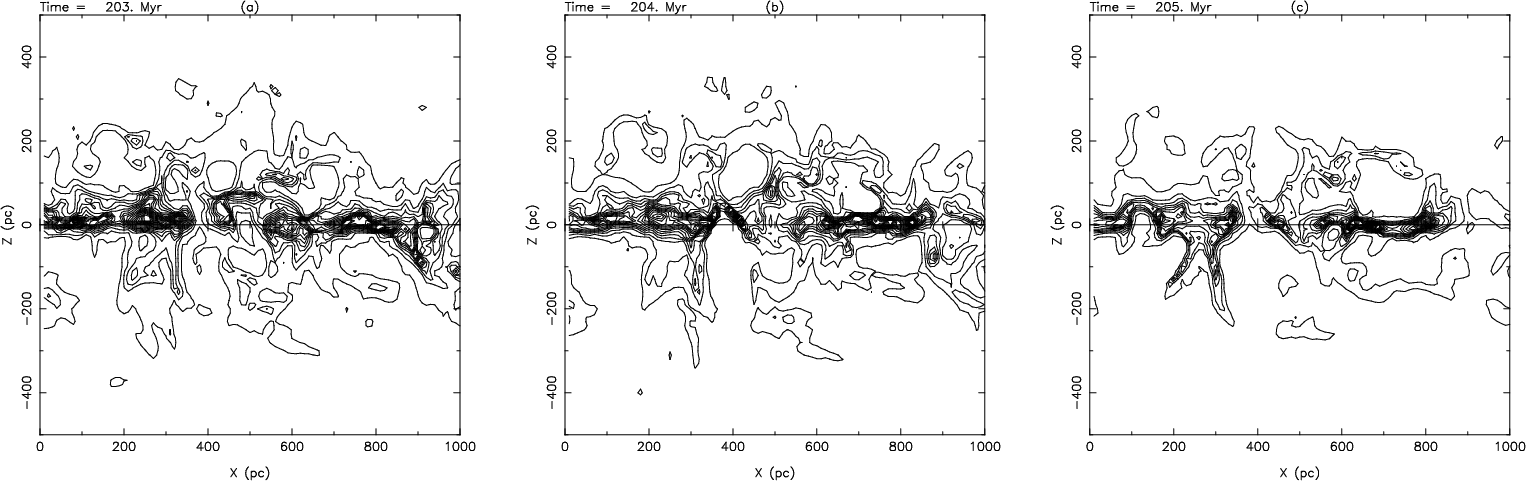,angle=-90,width=\hsize,clip=}
\caption{HI emission maps of the previous figure. The figure presents twenty equally spaced contours with column densities varying between $10^{18}$ and $10^{20}$ cm$^{-2}$ of the disk gas. One of the walls of the chimney is destroyed by the expansion of supernova located in its vicinity. The remaining wall has a thickness of several parsec and an elongation of 220 pc. At $(x=150, z=200)$ pc a shell is moving upwards (image (a)) detaching from other cold gas located at $(x=220, z=150)$ pc (image (b)) and later it break up into small clouds with sizes of several parsec (image (c) at $100 \leq x \leq 200, 150 \leq z\leq 240$ pc).}
\label{mavillez_fig17}
\end{figure*}
On average, after $5\times 10^{5}$ years, a type II supernova releases $295 \msolar$ of hot gas into the surrounding medium. Therefore, when several isolated supernovae merge together, they contribute to the formation of ``reservoirs'' of hot gas with enough energy to overcome the gravitational pull of the stellar disk expanding buoyantly upwards (Figures \ref{mavillez_fig3}, \ref{mavillez_fig6} - \ref{mavillez_fig9}). 

In its ascending motion, the hot gas interacts with the denser gas distributed in the thick disk. Such a configuration is Rayleigh-Taylor unstable and as a consequence, the hot gas acquires a finger-like structure with a mushroom cap. As it moves upwards it appears to carve the cooler layers that compose the thick disk (Figure \ref{mavillez_fig4}).

Bubbles and superbubbles occurring in the outer parts of the thin disk, at some $\sim 100$ pc above and below the Galactic plane, expand vertically due to the local stratification of the ISM becoming elliptical (Figures \ref{mavillez_fig6} and \ref{mavillez_fig8}). The vertical elongation of the shell surrounding the bubbles/superbubbles depends not only on the local medium, but also on the amount of energy released by the supernova explosions involved in the production of the bubble/superbubble.  

During their vertical expansion the shells, eventually become Rayleigh-Taylor unstable and blow their caps, releasing their hot inner parts into the ISM through narrow channels. The vertical expansion of the hot inner parts of the bubbles is halted by pressure effects due to the thick disk gas, while superbubbles break through it, blowing holes in the thick disk gas. The presence of such structures is shown in Figures \ref{mavillez_fig4} and \ref{mavillez_fig16}-\ref{mavillez_fig17}.

These structures may be classified as mini-chimneys and chimneys depending on their vertical elongation and on the presence of actual tunnels breaking through the disk gas as can be seen in Figure \ref{mavillez_fig4} (c) and (d). The tunnel observed in the figure crosses the thick gas disk and has a characteristic width of approximately 120 pc. The tunnel maintains the same appearance providing enough supernovae occur in the underlying association. 

The expansion of hot gas in well collimated structures may not have enough energy to break through the gas disk (Figure \ref{mavillez_fig4} (a) and (b)). Such a structure may be classified as a mini-chimney. The walls of the tunnels are formed by the remains of the broken shells whose appearance resembles worms crawling out of the disk. The average width of mini-chimneys varies between 50 and 60 pc whereas that of chimneys ranges from 100 pc to a few hundred parsec, depending on the number of supernova explosions inside the superbubble. 
\begin{figure*}
\centering{\hspace*{-0.1cm}}
\psfig{file=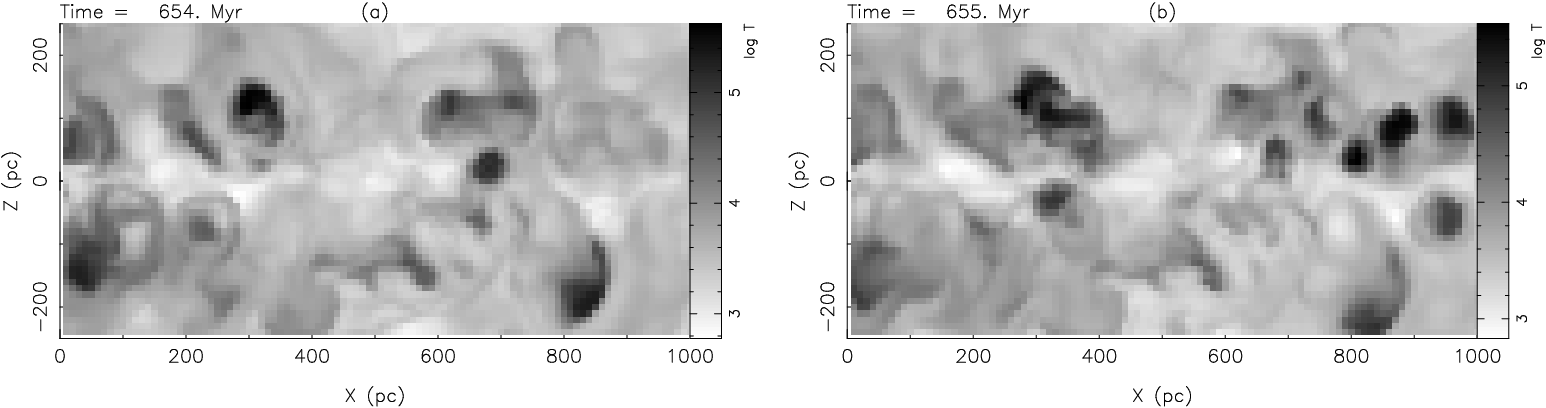,angle=-90,width=5.5in,clip=}
\psfig{file=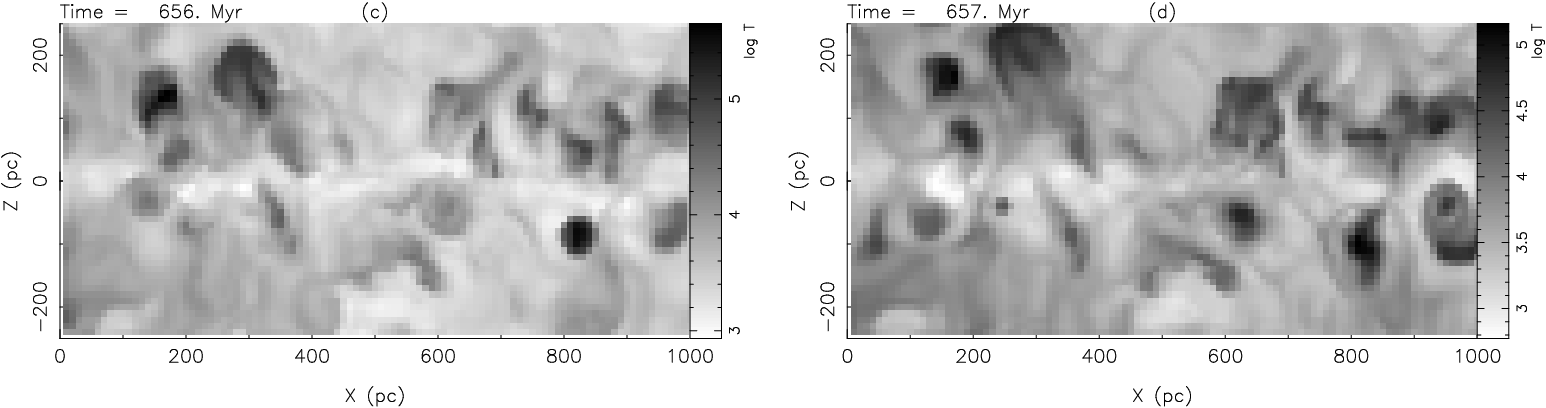,angle=-90,width=5.5in,clip=}
\caption{Time evolution of the temperature of the disk gas measured at $y=220$ between 654 and 657 Myr. The brightest regions are cold gas whereas the hottest regions are dark grey. The thin HI disk has a temperature of some $10^{2.8}=630$ K, whereas the hottest regions have temperatures of $\sim 10^{5.5}-10^{6}$ K. The thin HI disk has a temperature of some $10^{2.8}=630$ K and a thickness ranging from tens of pc to a hundred parsec or more in some locations. The hot region located at $(280 \leq x \leq 350, 70\leq z\leq 150)$ pc (image (a)) expands acquiring a mushroom-like structure as result of the interaction with the cooler medium above it. The mushroom is already formed at 655 Myr (image (b)) and acquires an increased elongation as it evolves (images (c) and (d)). A stream of hot gas is observed at the base of the mushroom which is connected to the underlying hot source.} 
\label{mavillez_fig18}
\centering{\hspace*{-0.1cm}}
\psfig{file=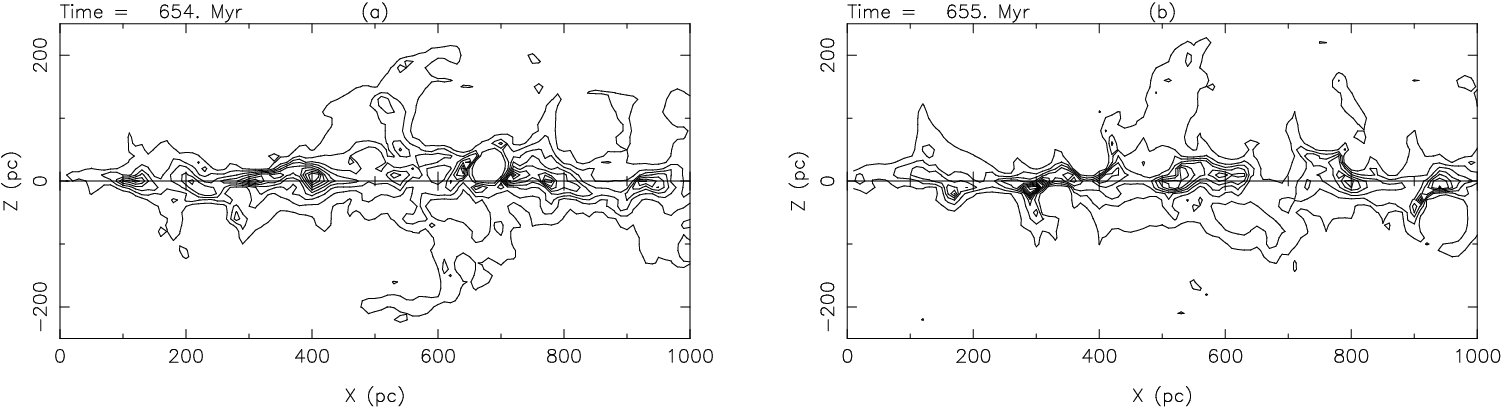,angle=-90,width=5.5in,clip=}
\psfig{file=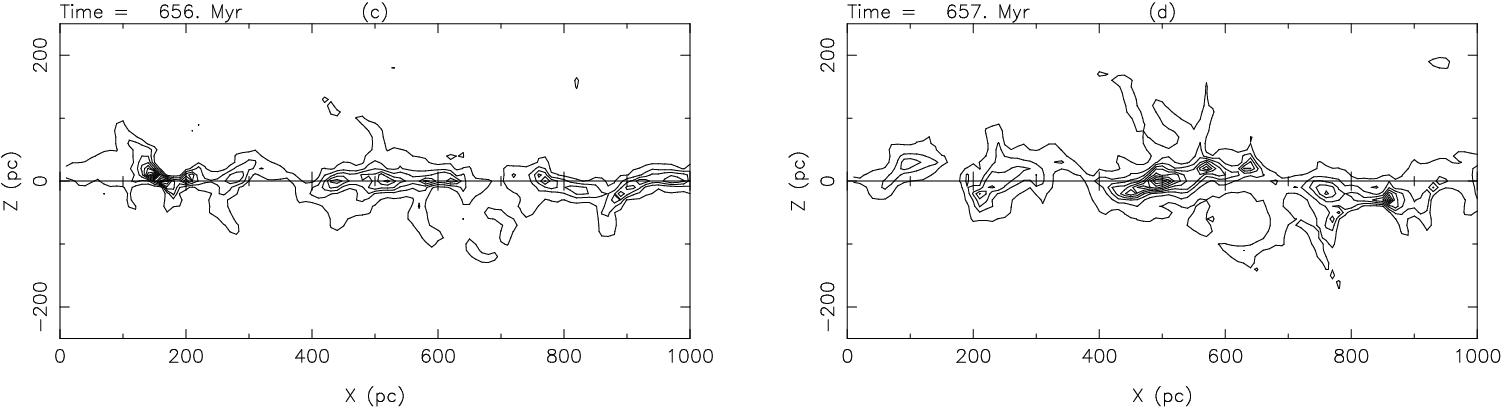,angle=-90,width=5.5in,clip=}
\caption{Emission maps of the disk gas shown in Figure \ref{mavillez_fig18}. The figure presents ten equally spaced contours representing column densities $\log(N_{HI})=18$ and $\log(N_{HI})=20$. The images show the presence of small HI clouds with a filamentary structure and connected to the underlying disk. Supernova cavities are surrounded by shells whose column density decreases with $z$.}
\label{mavillez_fig19}
\end{figure*}
Figure \ref{mavillez_fig16} shows the sequential evolution of the disk gas in a region where a chimney is formed, while Figure \ref{mavillez_fig17} presents the HI positional maps of the same region. The chimney, formed from a superbubble with a vertical elongation of 200 pc and a width of 100 pc, is located in the region $\sim 350\leq x\leq 450$ pc in the southern hemisphere (hereafter called region A). The base of the superbubble is located in the midplane and connects with two merging bubbles located in the northern hemisphere at $300\leq x\leq 480$ pc and $40\leq z\leq 150$ pc. After a million years, these two bubbles merge, creating a large cavity extending from $350\leq x\leq 480$ pc and $50\leq z\leq 190$ pc (Figures \ref{mavillez_fig16} and \ref{mavillez_fig17} (b)) and hereafter named region B.

At time 203 Myr, a supernova with radius of some 50 pc (the supernova is well inside phase III of its evolution) is observed to occur at $x=400$ pc, $z=-100$ pc, inside region A. similar explosions occurred in the superbubble leading to its expansion and break up of the top of its shell (Figure \ref{mavillez_fig16} (b). This is corroborated by the HI emission map in Figure \ref{mavillez_fig17} (b). A narrow tunnel of low density gas is surrounded by thick walls (thicknesses of a few parsec). The top of the chimney has a width of some 80-90 pc.

The hot inner parts of the superbubble are released upwards through sucessive injections due to different supernovae occurrences inside the cavity. As the gas expands out of the superbubble, it pushes upwards from $z=-300$ to $z=-450$ pc, debris of old shells in a period of 200 Myr. Continuing explosions inside the superbubble provoke the escaping of hot gas not only in the southern hemisphere, but also inject gas into region B (Figure \ref{mavillez_fig16} (c)). As a consequence the upper parts of the shell surrounding region B also breaks up and an injection of hot gas into the thick disk through a mini-chimney is induced. 

The gas released by the  mini-chimney is unable to break through the thick disk and acquires a finger-like configuration until it cools down to a recombination temperature and merges together with the surrounding medium (Figure \ref{mavillez_fig17}). The figure shows the evolution of the same region in space as Figures \ref{mavillez_fig16} and \ref{mavillez_fig17}, but at later times: 209-211 Myr. After a period of a few million years, the width of region A varies between 200 pc near the plane to 110 pc on the upper parts of the shell (Figure \ref{mavillez_fig17} (c)).

\subsection{Cold Gas in the Disk}

HI emission maps showing the vertical distribution of the cold gas in the disk and the corresponding temperature maps, taken at along the plane crossing the midplane at $y=220$ pc, are presented Figures \ref{mavillez_fig18}-\ref{mavillez_fig19}. The figures present a time evolution of the cold gas in the disk during a period of four million years and show 20 equally spaced contours representing values of column density between $\log(N_{HI})=18$ and $\log(N_{HI})=20$, where $N_{HI}$ is in cm$^{-2}$. The concentration of contours in a specific direction indicates a large column density of the gas there.
\begin{figure*}
\centering
\psfig{file=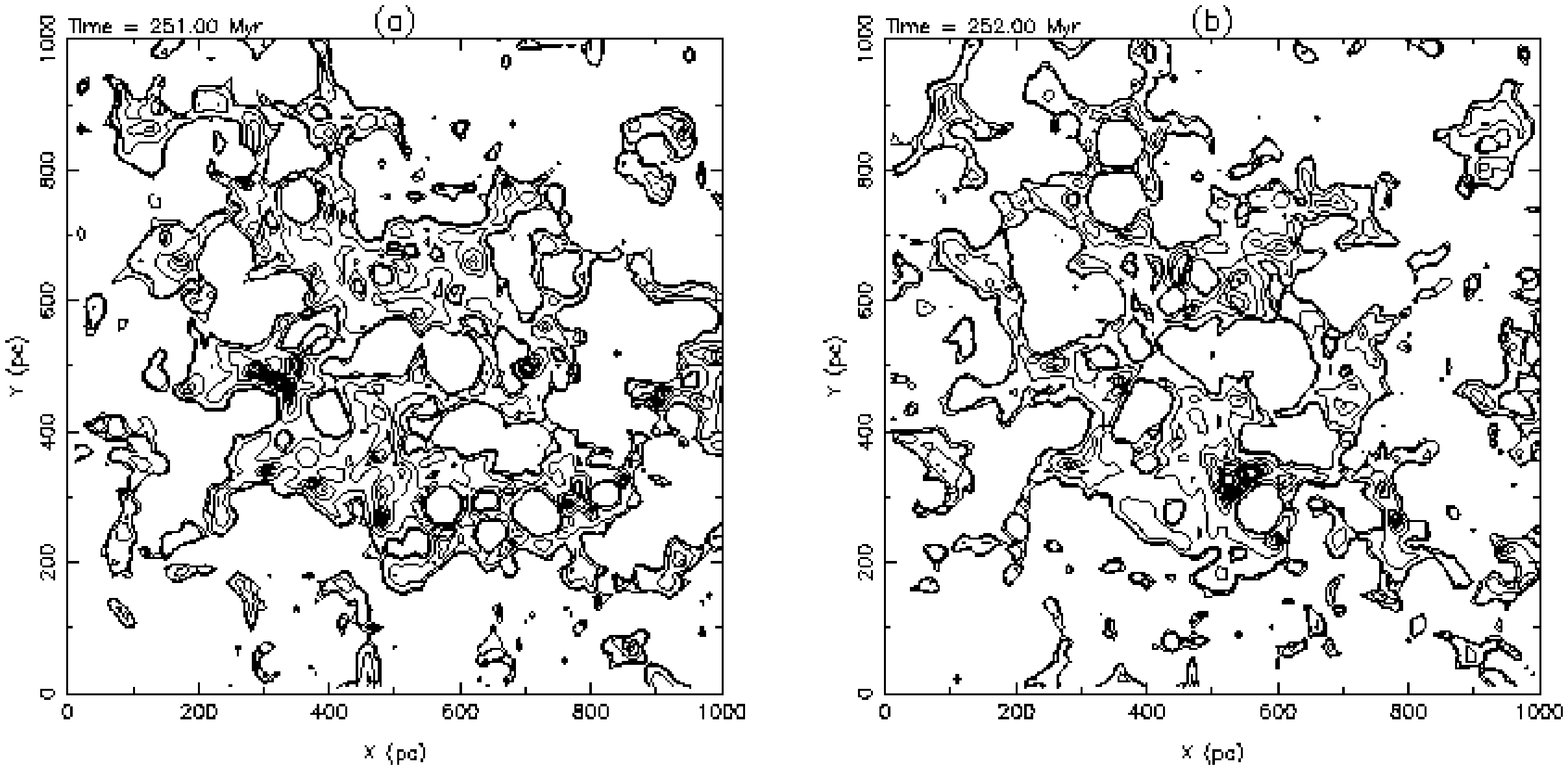,angle=-90,width=5.in,clip=}
\psfig{file=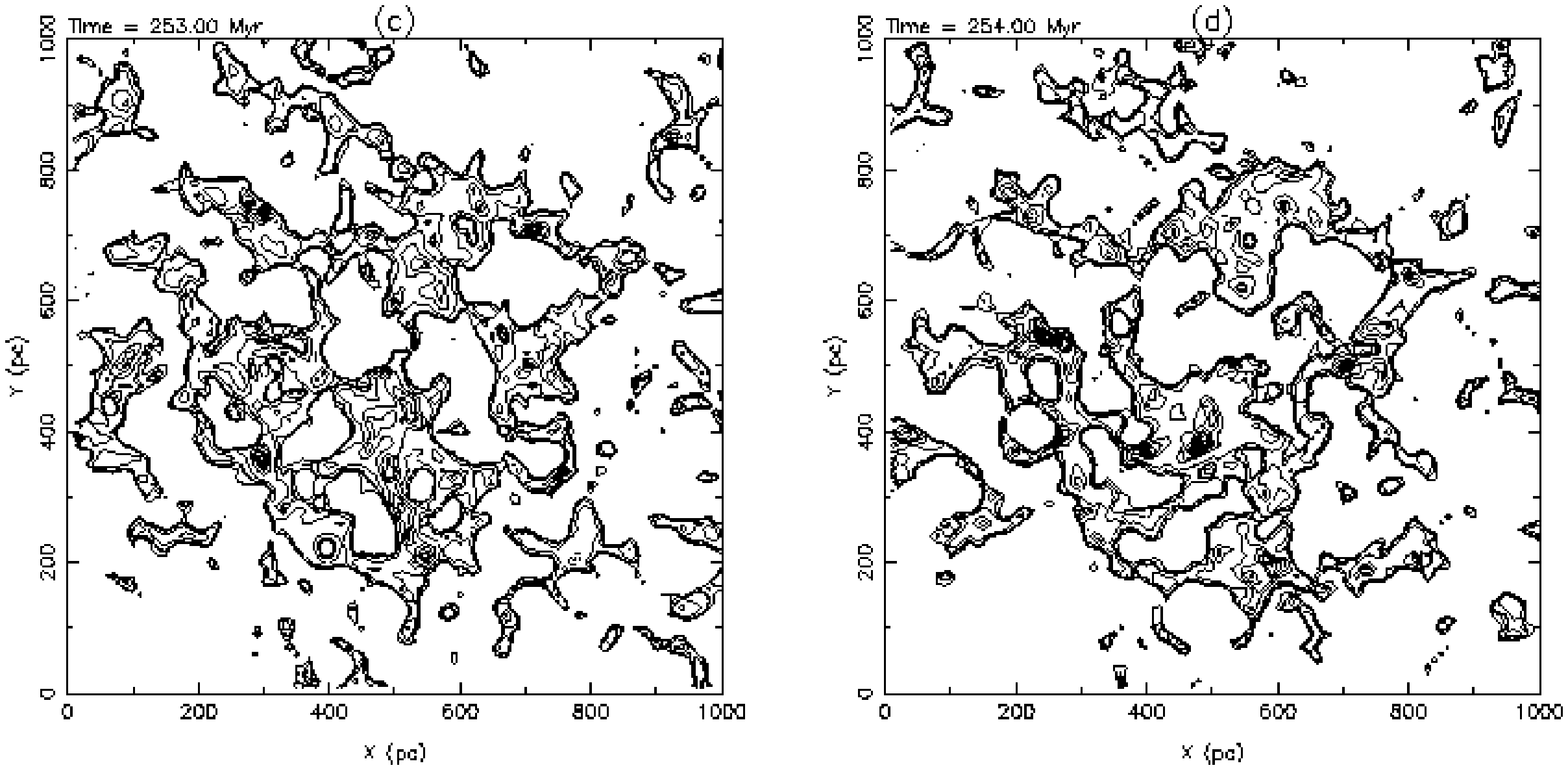,angle=-90,width=5.in,clip=}
\caption{HI emission of the cold gas at z=0 pc and taken at:(a) 251 Myr, (b) 252 Myr, (c) 253 Myr and (d) 254 Myr of simulated time. The images show the time evolution of the cold gas in the midplane. The images show a sheet of cold gas with holes and channels of low density material forming a network of hot gas. Small clouds with sizes of few parsec result from the break up of larger structures. The presence of holes is a transient phenomenon that depends on the location of new supernovae as well as from the time of cooling of the disk gas.}
\label{mavillez_fig22}
\end{figure*}

In general the figures show HI emission concentrated in a wiggly cold structure (thin HI disk) distributed along the Galactic plane, having a thickness of a few tens of parsec, and in some cases being as thin as 10-20 pc. The presence of a very thin HI disk (with $\Delta z\sim 10-20$ pc) is an occasional phenomenon and results from local variations in the structure of the thin disk as result of, for example, correlated supernovae located near the plane or collisions of high velocity clouds with the thin disk. The occurrence of the supernovae leads to the formation of tunnels of hot gas crossing the Galactic plane, and therefore, to the absence of the thin disk observed in some of the $xz$-maps (see Figure 9). Collisions of high velocity clouds, may compress and drag with the cloud parts of the thin disk. The largest column density observed in the images is caused by the low scale height of the cold gas.

The figures show the presence of low column density regions (absence of contours) persisting for long periods. These regions result from supernovae that occurred some time in the past, and that have blown holes in the Galactic disk. As the expanding gas cools, a thin shell of colder gas is formed behind the blast-wave generated by the supernova. 

The vertical structure of the HI holes is observed in most of the figures. Some are partially surrounded by thin sheets of cold gas, which result from old shells. The cone-like structures of low column density surrounded by cold sheets are found located on each side of the Galactic plane. Because of the high column density in the plane, the cones slightly change its local structure, but are unable to break the thin disk.

Shells displaced from the midplane have a non-spherical appearance, resulting from the local medium not being smooth (on the small scale - 5 pc) and partly from the steep variation of the density along the $z-$direction - the shells acquire a ovoid form (Figures \ref{mavillez_fig17}, \ref{mavillez_fig18} - \ref{mavillez_fig19}). As the bubbles expand upwards the shells break, releasing the hot inner parts of the remnant into the surrounding medium. Therefore, the remaining parts of the shells become the walls of the cone-like structures. These walls have a wiggly appearance resembling ``worms'' crawling into the halo.
\begin{figure*}
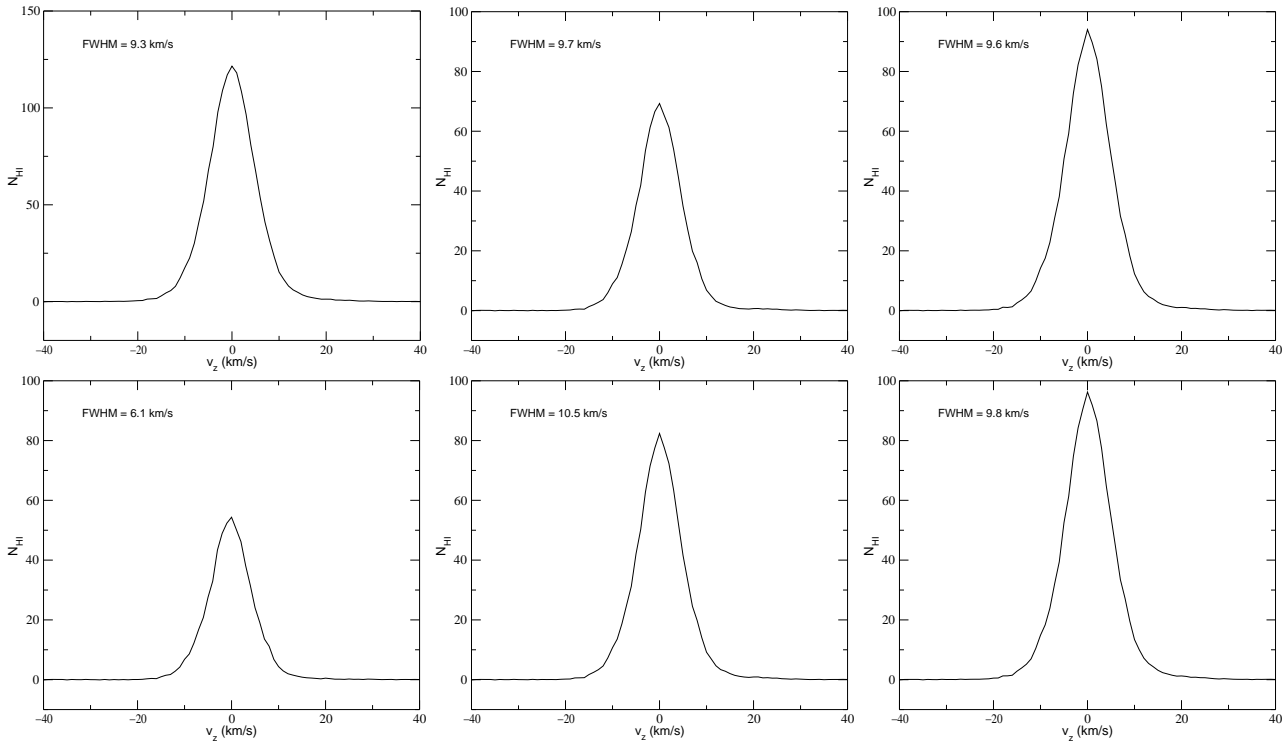

\centering{}
\psfig{file=mavillez_fig17-a.eps,angle=0,width=2.2in,clip=}\hspace*{0.1cm}\psfig{file=mavillez_fig17-b.eps,angle=0,width=2.2in,clip=}\hspace*{0.1cm}\psfig{file=mavillez_fig17-c.eps,angle=0,width=2.2in,clip=}\\
\psfig{file=mavillez_fig17-d.eps,angle=0,width=2.2in,clip=}\hspace*{0.1cm}\psfig{file=mavillez_fig17-e.eps,angle=0,width=2.2in,clip=}\hspace*{0.1cm}\psfig{file=mavillez_fig17-f.eps,angle=0,width=2.2in,clip=}
\caption{Column density (N$_{HI}/10^{18}$ cm$^{-2}$) of cold gas in the disk as a function of the velocity (km s$^{-1}$) along one column (taken at $-100\leq z\leq 100$ pc). The extreme values of the FWHM are 6.1 km/s and 10.5 km/s with the average values varying between 9.3 and 9.8 km/s.}
\label{mavillez_fig23}
\end{figure*}

Face-on maps of HI emission from the midplane, taken in a six million years period (Figure \ref{mavillez_fig22}), show high column density regions spread over the plane surrounding holes and channels of low column density gas. The holes and tunnels form a large scale network of low density gas spread over the Galactic disk. The tunnels have widths of some 100 pc and lengths in the range of 200-400 pc. This network forms as a result of young supernovae occurring near to old remnants. High velocity shock waves sweep up the old remnants, disrupting their shells and leading to a movement of the hot gas inside the young remnants into the low density medium surrounding it. The blast wave may suffer several reflections inside the tunnel, leading to the formation of small clouds as result of the disruption of the shells and condensation into clouds of the gas swept several times by the blast waves. This gas is compressed during each swept up to four times the local density (assuming that these shocks are strong). The increase in the density leads to an increase in the cooling of the material and therefore to its condensation into clouds.

Most of the gas observed in the HI emission maps forms clouds with sheet like appearance and connected to each other. Clouds have sizes/thicknesses varying between few parsec to hundreds of parsec. The larger clouds suffer Rayleigh-Taylor instabilities and break up into sheet-like structures that fall towards the Galactic plane. However, isolated structures with large sizes, identified near the plane are moving towards it colliding with it. The collision leads to the compression of the thin disk where higher density gas accumulates forming possible giant molecular clouds where star formation would occur. 

Column density profiles of the HI gas along a column of 400 pc (with $-200\leq z\leq 200$ pc) versus velocity are presented in Figure \ref{mavillez_fig23}. The figure presents profiles convolved with a Gaussian that has a velocity dispersion that corresponds to the local gas temperature, being each cell of the column normalized to the local gas density. The profiles are symmetric with respect to the central peak and have FWHM values varying between 6.1 and 10.5 km s$^{-1}$, being these the extreme cases. The average values of FWHM are $\sim 9.5$ km/s. 

The major fraction of cold gas in the disk has positive and negative velocities varying between $-20$ and $20$~km/s. However, there are clouds with larger velocities indicating a fast approach to the midplane, where they collide with cold gas falling from the opposite hemisphere, or an escape from the thin disk (positive velocities). 

\section{Discussion}
\begin{figure*}
\mbox{\epsfxsize=5.5in\epsfysize=4in\epsfbox{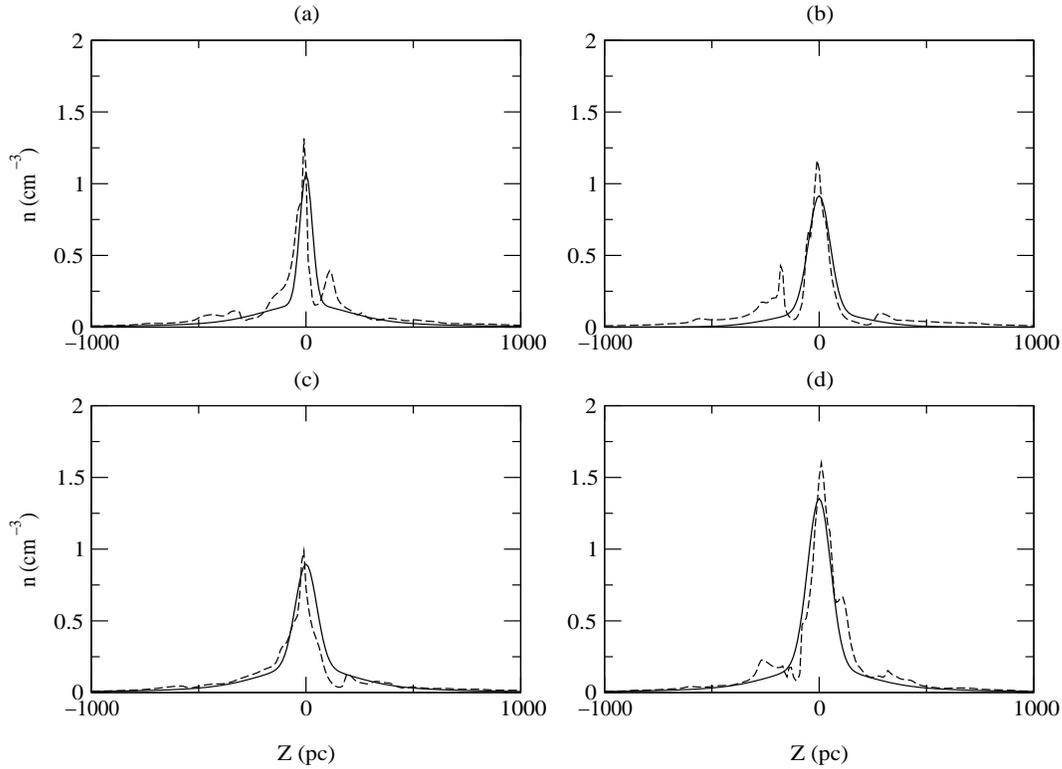}}
\caption{Best fits (solid line) for $n(HI)$ distribution in $z$ (dashed line) along four lines of sight crossing the disk plane at: (a) $(x=30, y=30)$ pc, (b) $(x=300, y=250)$ pc, (c) $(x=600, y=500)$ pc and (d) $(x=800, y=800)$ pc. Note the presence of humps in the $n(z)$ distribution on each side of the thin disk. These humps prevent the hot gas coming from the thin disk to expand freely into the thick disk.}
\label{mavillez_fig24}
\end{figure*}
The results of large-scale modelling of a section of a Galactic arm, located at 8.5 kpc from the Galactic centre are presented in this paper. It is a three-dimensional model that assumes the ISM as an integrated system where the collective effects of supernovae (isolated and clustered) are taken into account. The supernovae occur in a medium strongly structured by previous supernovae, ionization fronts and radiative cooling. 

The model considers the interstellar gas distributed in a smooth thin disk with the vertical distribution of the cool and warm neutral HI gas given by Dickey \& Lockman (1990) and warm ionized gas given by Reynolds (1987). It expli-\\citly includes both correlated and random supernovae in a manner compatible with observations. Sixty percent of the supernovae occur within associations, whereas the remaining occur isolated. The supernovae belong to types Ib, Ic and II, and their masses are taken into account, as mass loading is allowed. 

Once disrupted by the explosions, the disk never returns to its initial state. Instead, it approaches a state where a thin HI disk is formed in the Galactic plane, overlayed by a thick disk of warm gas. 

The thick disk is generated as long as enough supernovae occur in the disk, regardless of the initial density profile adopted for the disk gas (\S 3.1), and, therefore, its formation and stability are directly correlated to the supernova rate per unit area in the simulated disk. 
\begin{table*}
\begin{tabular}{cccccccc} 
\hline
Distribution & A & B & C & D & E & F & H$_{z}$ \\
               & (cm$^{-3}$)  & (pc) & (cm$^{-3}$)  & (pc) & (cm$^{-3}$) & (pc) & (pc) \\
\hline
\hline
DL & 0.395 & 90 & 0.107 & 225 & 0.064 & 403 & 177.6\\
Fig. \ref{mavillez_fig24} (a) & 0.90 & 30 & 0.110 & 200 & 0.07 & 405 & 158.6\\
Fig. \ref{mavillez_fig24} (b) & 0.81 & 50 & 0.101 & 200 & 0.04 & 405 & 191.7\\
Fig. \ref{mavillez_fig24} (c) & 0.70 & 50 & 0.105 & 205 & 0.08 & 405 & 182.4\\
Fig. \ref{mavillez_fig24} (d) & 0.76 & 54 & 0.105 & 208 & 0.09 & 405 & 203.6\\
\hline
\end{tabular}
\caption{Values of the coefficients $A, B, C, D, E$ and $F$ in equation (\ref{den1}) and effective scale height H$_{z}$ for the density profiles of Dickey \& Lockman (1990) and of Figure \ref{mavillez_fig24} (a)-(d).}
\end{table*}
The density distribution of the disk gas in $z$ can be approximated by a linear combination of two gaussians and an exponential with the general form, based on the model of Dickey \& Lockman (1990)
\begin{equation}
\label{den1}
n(z)=\frac{n_{\circ}}{0.566}\left[A e^{-Z^{2}/2B^{2}}+C e^{-Z^{2}/2D^{2}}+Ee^{-Z/F}\right]
\end{equation}
where $n$ is the number density of the ISM phase, $n_{\circ}$ is the number density at the Galactic plane, $A$, $C$ and $E$ are paremeters written in units cm$^{-3}$ and $B$, $D$ and $F$ are written in pc. The density distribution yields an effective scale-height
\begin{equation}
\label{den2}
H_{z}=\frac{1}{n_{\circ}}\int_{0}^{+\infty}n(z) dz.
\end{equation}
Noting that 
\begin{equation}
\int_{0}^{+\infty} e^{-Z^{2}}dZ=\frac{\sqrt{\pi}}{2},
\end{equation}
and using (\ref{den2}) the effective scale-height of the density distribution in equation (\ref{den1}) is given by
\begin{equation}
\label{scale}
H_{z}=\frac{1}{0.566}\left[(AB+CD)\sqrt{\frac{\pi}{2}}+EF\right].
\end{equation}
For the density distribution of the disk gas in the Galaxy, Dickey \& Lockman (1990, hereafter denoted by the DL distribution) inferred the parameters shown in the first row of Table 3. The table presents, in rows 2-5, the parameters corresponding to density distributions measured along four lines of sight crossing the disk at: (a) $x=30$, $y=30$ pc, (b) $x=300$, $y=250$ pc, (c) $x=600$, $y=500$ pc and (d) $x=800$, $y=800$ pc. These distributions are the best fits to the underlying curves of the density distribution as shown in Figure \ref{mavillez_fig24}.

As can be seen in Table 3, the coefficients $C, D, E$ and $F$ are similar between all the fits and the DL distribution, whereas the $A$ and $B$ parameters for the fits and for the DL distribution have variations as much as $15\%$. The difference in the extreme case indicates that, in the simulations, most of the cold gas in the disk is concentrated near the plane and thus has a smaller effective scale-height.

Despite the differences between the DL distribution and those obtained in the simulations, the thick gas disk has a distribution compatible with the presence of two gas components having different scale heights: a neutral layer with a scale height of 500 pc (warm HI disk) and an ionized component extending to a height of 1 to 1.5 kpc above the thin HI disk. 

Similar distribution of the thick gas disk and the HI and HII thick disks has been observed in the Galaxy (Lockman, 1984; Lockman {\rm et~al.}, 1986; Reynolds 1987) and in external galaxies such as NGC 891 (Dettmar, 1992). Studies of the z-distribution of HI in edge-on galaxies (NGC 4565 and NGC 891) carried out by Rupen (1991) have shown the presence of an HI layer around the galactic midplane. This layer has a thickness that varies from 120 pc to 300 pc in both galaxies. 

The average physical properties of the different phases in the simulated disk in the steady state evolution are summarized in Table 3. These values are in agreement with those derived by Karberla \& Kerp (1999) for the best-fit parameters required for a statistical steady equilibrium of the halo and disk gas in the Galaxy.
\begin{table}
\begin{center}
\begin{tabular}{lccc}
\hline
Component & $n$ & T & H$_{z}$\\
& (cm$^{-3}$) &  (K) & (kpc)\\
\hline 
\hline
Thin HI disk & $0.6-1$ & $\leq 8\times 10^{3}$ & 0.09 - 0.12 \\
Thick HI disk & $0.1$ & $10^{4}$ & 0.5 \\
Thick HII disk & $0.01-0.02$ & $10^{4} - 10^{5}$ & 1 - 1.5\\
\hline
\end{tabular}
\caption{The table presents the physical properties of the thin and thick disk gas in the simulated volume. $n$, T and H$_{z}$ denote the volume density, temperature and height of the disk gas.}
\end{center}
\end{table}
\subsection{Volume Filling Factors of ISM Phases}
The volume filling factors of the different ISM phases have been presented in \S 3.1 and the corresponding best-fit profiles are presented in Figure \ref{mavillez_fig3}. The distribution of the hot gas varies from a occupation volume of some $23\%$ in the inner disk (below 500 pc) to $100 \%$ above 2.5 kpc. 

The thin disk is populated by a cold medium with a volume-filling factor smaller than $5 \%$. The warm neutral gas dominates below 500 pc occupying a volume of some $50\%$, whereas the warm ionized medium dominates between 500 and $\sim 1500$ pc. At some 1500 pc, both the warm ionized and hot medium fill the same volume. This region corresponds to the disk-halo interface described above. In what follows, we shall consider the disk-halo interface as the height at which the two media have the same volume-filling factor. 

As the warm neutral and ionized media dominate\\different regions, it gives the appearance that each of these phases constitute different layers. However, both media are mixed and have different scale-heights. The region dominated by the ionized medium is known as the Reynold's layer (Reynolds, 1987).

The filling factors of the different phases observed in this study are similar to those estimated by Spitzer (1990) and  Ferri\`ere (1995). These authors estimated a volume-filling factor of the hot gas in the Galactic disk of $\sim 0.2$. In addition, Ferri\`ere (1995) estimated volume-filling factors for the cold and neutral gases in the disk similar to those shown in Figure \ref{mavillez_fig3} and discussed above. 

Ferri\`ere (1998) estimated the variation of the volume-filling factor of the hot gas with $z$. She finds that, at the solar circle, the volume occupation of the hot gas increases from 0.15 at $z=0$ pc to 0.23 at $z=200$ pc and decreases gradually for $z>200$ pc (see Figure 12 in Ferri\`ere, 1998). The latter result seems to be in conflict with the calculations reported in this paper. Such apparent conflict is resolved if one considers that the calculations carried out by Ferri\`ere reflect the distribution of hot gas inside cavities of supernova remnants and superbubbles, whose distribution decreases with $z$, while the present paper also considers hot gas that bouyantly rose from interacting supershells. Ferri\`ere's model does not include that gas at all, and so does not model the high-$z$ region fully.

\subsection{Large Scale Outflows and Local Outbursts}
The simulations show that the thick gas disk is fed with the hot gas coming from the thin disk through two major processes: large scale outflows and local outbursts (known as chimneys).

Large scale outflows are the result of the buoyant expansion of hot disk gas concentrated in large reservoirs located on either side of the thin HI disk (as shown in the density profiles). These reservoirs are formed by the hot inner parts of isolated supernovae randomly distributed in the stellar disk. 

Such hot gas has enough energy to be held gravitationally and expands buoyantly upwards, generating large scale outflows (Figure \ref{mavillez_fig4}). The ascending flow triggers the growth of Rayleigh-Taylor instabilities as it interacts with the cooler, denser medium in the thick disk. Therefore, the hot gas acquires a finger-like structure with a mushroom cap on the tip of the finger. During the expansion of the finger, further instabilities develop, until the major fraction of the ascending gas cools and merges with the surrounding medium.

The gas in the finger cools from the outside towards the centre, thus giving it the appearance of being enveloped by a thin sheet of cooler gas (Figures \ref{mavillez_fig4} and \ref{mavillez_fig17} at $x=300,~y=100$ pc). A structure called the ``anchor'', with properties similar to those described above, has been observed in the southern Galactic hemisphere and reported by Normandeau \& Basu (1998).

Chimneys result from correlated supernovae, which generate superbubbles that acquire an elongated shape due to the stratification of the ISM in the $z-$direction. Superbubbles are able to expand and blow holes in the disk, injecting high speed gas, which forces its way out through relatively narrow channels with widths of some 100 - 150 pc. They provide a connection between the thin disk and the upper parts of the Reynolds layer.

One can identify mini-chimneys and chimneys from the break up of shells and supershells located at the upper parts of the thin disk. Mini-chimneys result from isolated supernovae that have had their shells broken and released to hot gas into the surrounding medium (the broken shell acts as the walls of a small tunnel produced by the blow out), whereas chimneys are large tunnels of hot gas crossing the thick HI disk. These structures are well collimated, with a maximum width varying between 50 pc for mini-chimneys and 150 pc for chimneys, and vertical extensions varying between few tens of parsec (mini-chimneys) and 500 pc or more for the chimneys. 

This is corroborated by the HI position maps, taken at different locations of the disk along lines of sight parallel to the plane, showing the presence of broken shells (``worms'') wiggling around in the $z$-direction. Face-on HI maps of the disk taken at $\left|z\right|\sim 250$ pc show the presence of low density regions with elliptical or circular forms and having a large range of sizes. The major fraction of these holes have been produced by large cavities of hot gas resulting from clustered supernovae, and therefore by chimneys. It is hard to identify the holes resulting from mini-chimneys, suggesting that their maximum elongation is some 150 pc in the $z$-direction.

Surveys of the Galactic disk (Heiles, 1984, 1990) and of M31 (Brinks \& Bajaja, 1986) and M33 (Deul \& Hartog, 1990) have revealed the presence of roughly elliptical holes in the HI maps, with sizes varying between 40 and 1000 pc. Some holes show a clear shell structure, although for most holes the shells are not resolved. 

The majority of the shells have deviations in the local velocity field with expansion velocities of 10-30 km s$^{-1}$. The estimated kinematic ages of these holes varies from 2.6 to 30 Myr and the energy requirements are $10^{49}$ to $10^{53}$ erg. A correlation between the holes in the HI maps with single HII regions and OB associations in these galaxies has been established by Deul \& Hartog (1990).

Heiles (1984) identified the presence of holes in the HI emission maps of the Galactic disk. In addition, sheets of HI gas with a vertical structure have been identified in these maps by Heiles and Koo {\it et al.} (1992). These vertical structures have been associated with broken shells and supershells in the Galaxy, and have been proposed as being the walls of chimneys by Tomisaka \& Ikeuchi (1986) and Norman \& Ikeuchi (1989). HI maps of NGC 891 show a series of low density channels, similar to the ones shown in this paper.

Most recently, Normandeau {\it et al.} (1996) identified a superbubble in the Perseus arm underlying an association of nine O-type stars. The superbubble has been created by stellar winds from these stars (Basu {\it et al.}, 1999) - no supernova has been identified within the W4 region - and has an elongated appearance resulting from the vertical stratification of the disk gas. Although in HI emission maps the region looks like a chimney, such a view is altered in the H$\alpha$ emission maps of the region, which clearly shows the presence of a superbubble. 

\subsection{Numerical Resolution}
Numerical resolution seems to be an issue in the simulations reported in this paper and in Paper II. The major constraint imposed by numerical resolution is the prevention of the formation of very small scale structures resulting from thermal instabilities and condensations due to radiative cooling. Furthermore, the amount of cooling may be substantially increased if the resolution is great enough to allow for the development of turbulent shear layers in regions of outflow and infall. 

The numerical approach adopted to minimize such problems combined the usage of a high-resolution scheme (the PPM) to advance the solution in time with the dynamical refinement of the computational grid whenever required. The latter was set up in the region located at $-150\leq z\leq 250$ pc (hereafter referred to disk-grid), whereas in the rest of the computational domain no dynamical refinement was used - the maximum resolution being 10 pc. The resolutions in the disk-grid were 5, 2.5 and 1.25 pc (this was the smallest grid size allowed during the calculations).

The PPM is a third-order Godunov scheme, which does not rely on artificial viscosity to spread sharp variations of the flow variables. These discontinuities are represented by sharp profiles, free of spurious oscillations, of the flow variables and are treated as solutions to the Riemann problem.

\section{Future Efforts and Final Remarks}
The major drawback of the simulations is the absence of magnetic fields that may constrain the upward motions of the disk gas. Understanding the disk-halo interaction requires simulations that couple together the major consituents of the ISM in the disk and halo, i.e, gas, magnetic fields, heating and cooling. 

The presence of magnetic fields may provide a confinement effect over the disk gas, providing the scale height of its distribution is much larger than that of the disk gas. For an average density distribution such as that observed in the simulations, the scale height is of the order of $H_{g}\sim 170$ pc, which is comparable to the scale height of the thin component of the Galactic magnetic field - $H_{B}\sim 170$ pc (Thompson \& Nelson, 1980). Tomisaka (1998), using a magnetic field decreasing as the square root of the gas density, shows that no confinement occurs for the superbubbles in the disk. However, the presence of a thicker component of the magnetic field, with a scale height of 1.2 kpc (Han \& Qiao, 1994), confines the expansion of superbubbles, located at $\left|z\right|\leq 300$ pc, to more than 20 Myr. 

Furthermore, as has been shown here, such study must be three-dimensional and independent of the grid size. That is, a 3D grid with a thickness of only a few hundred parsec generates an overpressured region in the disk, and thus the simulations evolve in a two-dimensional fashion rather than three-dimensional. The time delay between supernovae and superbubbles is so large that when the next supernova explodes, the previous one has already expanded to few hundred parsec in size, and therefore crossed the boundaries of the grid. The younger supernova has a large probability of occurring inside the old cavity and, as such, it expands in an evacuated medium with velocities larger than that of a supernova in phase II. 

Such simulations are difficult to carry out due to present-day computational facilities. However, efforts are under way, by the author, to provide a new model where magnetic fields come into play in a medium where the disk-halo structure is regulated by supernovae. 

Phenomena such as high and intermediate velocity clouds are regarded as products of the Galactic fountain and their dynamics, structure and evolution are described in paper II (Avillez, 1999b). In addition, some of the items described in this paper require full discussion which will be presented in future papers in the series.
\section*{Acknowledgments}
This paper is dedicated to the memory of Prof. Franz D. Kahn. He will be sadly missed. 
I would like to thank the anonymous referee whose comments and suggestions led to the improvement of this manuscript.

\label{lastpage}
\end{document}